\documentclass[aps,prl,reprint,showpacs,showkeys,superscriptaddress,preprintnumbers,amsmath,amssymb,nobibnotes, nofootinbib]{revtex4-2}
\usepackage{multirow}
\usepackage{graphicx}
\usepackage{bm}
\usepackage{dcolumn}
\usepackage{picture}
\usepackage[hidelinks]{hyperref}
\usepackage{natbib}
\usepackage{amsmath}
\usepackage{times}
\usepackage{color}
\usepackage{xcolor}
\usepackage{soul}
\usepackage{etoolbox}
\usepackage{graphicx}
\usepackage{array}
\usepackage{rotating}
\usepackage{amsmath}
\usepackage{amssymb}

\begin{document}
	\title{Protracting the Weyl phase by a giant negative lattice expansion in Bi doped Sm$_2$Ir$_2$O$_7$}
	\author{Prachi Telang}
	\affiliation{Department of Physics, Indian Institute of Science Education and Research, Pune, Maharashtra-411008, India}
	\author{Surjeet Singh}
	\email[email:]{surjeet.singh@iiserpune.ac.in}
	\affiliation{Department of Physics, Indian Institute of Science Education and Research, Pune, Maharashtra-411008, India}

	%\affiliation{Department of Physics, Indian Institute of Science Education and Research, Pune, Maharashtra-411008, India}\affiliation{Center for Energy Science, Indian Institute of Science Education and Research, Pune, Maharashtra-411008, India}

\date{\today}

\begin{abstract}
We show that the Weyl phase in $\rm Sm_2Ir_2O_7$ is protracted up to at least  2~\% alloying with Bi by an anomalous negative lattice expansion (NLE) with $\rm \Delta a \sim- 0.01~\AA$. With further doping, the magnetic ordering disappears and electrical resistivity decreases by orders of magnitude; the resistivity upturn remains but with $\rm 1/T$ dependence of Weyl phase changed to $\rm -lnT$ dependence characteristic of the Quadratic Band Touching (QBT). At the Weyl-QBT phase boundary, a new phase is evidenced whose resistivity scales as $\rm -T^{1/4}$ possibly due to proximity to a quantum critical point proposed several years ago [Phys. Rev. X 4, 041027 (2014)], but whose  experimental evidence has remained elusive thus far.      
\end{abstract}


\maketitle

%\section{Introduction}
%\label{Intro}
\textit{Introduction} - The study of topological phases of matter has emerged as one of the key research areas in contemporary condensed matter physics and material science. While the topological insulators having linearly dispersing electronic surface states came first to the fore, the current focus has shifted to topological semimetals where the symmetry protected topological states exist within the bulk of a material \cite{Senthil2015, LvBQ2021}. In last few years, a growing list of topological semimetals, including Dirac and Weyl semimetals, and a related nodal line semimetal, among others have been proposed. The Weyl semimetal (WSM) phase, that concerns us here, requires breaking of either the time-reversal symmetry (TRS) or the lattice inversion symmetry. The pyrochlore iridates are the prime candidates for realizing the TRS breaking WSM and other novel topological phases~\cite{Wan, Krempa2014}. In these pyrochlores, the all-in/all-out (AIAO) noncollinear magnetic order, which sets in upon cooling, breaks the time-reversal symmetry (TRS) while the lattice inversion symmetry remains intact. This lifts the degeneracy of the quadratic bands touching at the $\rm \Gamma$ point, leading to several pairs of Weyl nodes at points $\rm \pm \textbf{k}_i$~\cite{Wan}. Given these settings, it is interesting to investigate how the "topological protection" enjoyed by the Weyl phase would manifest itself if while preserving the lattice inversion symmetry, TRS is also restored by tuning an external perturbation in a controlled manner. While this question is of significant general interest in the study of topological phases, in the particular case of pyrochlore iridates it also interesting for another reason as Savary et al.~\cite{Lucile2014} showed the existence of a quantum critical point (QCP) at the interface of WSM and QBT (quadratic band touching). However, experimental evidence of the QCP or the associated quantum phase has been lacking till date.  
  
Here, we address this question by alloying $\rm Sm_2Ir_2O_7$ with Bi. Both $\rm Sm_2Ir_2O_7$ and $\rm Bi_2Ir_2O_7$ have the same pyrochlore structure. Their physical properties are well investigated in previous studies: while $\rm Bi_2Ir_2O_7$ is a correlated metal~\cite{Wang}, $\rm Sm_2Ir_2O_7$ is a candidate WSM with concomitant AIAO ordering and a metal-to-insulator transition (MIT) near 120 K~\cite{Wan, Donnerer}. Several experimental evidences of topological phases, including the WSM phase, in the pyrochlores iridiates have been furnished recently using  various experimental probes including optical conductivity~\cite{Sushkov2015, Ueda2016b}, electrical resistivity and magnetotrasnport~\cite{Tafti, Telang2019, UedaNC2018, UedaNC2017, Nakayama2016, Ueda2020}, and angle-resolved photoemission spectroscopy (ARPES)~\cite{Kondo2015} \par

In this letter, we show that the Weyl phase in $\rm Sm_2Ir_2O_7$ is protracted up to as far as 2\% Bi doping by a giant anomalous negative lattice expansion (NLE). In the Weyl phase up to $\rm x = 0.02$, the low temperature resistivity ($\rho$) is shown to follow a $\rm 1/T$ dependence as expected theoretically. With further doping, $\rho$ decreases by more than two orders of magnitude and simultaneously the AIAO ordering melts away marking the end of the Weyl phase and emergence of a semimetallic phase with $\rm \rho \propto -lnT$: A characteristic of the quadratic band touching (QBT). At the WSM-QBT boundary, a new phase has been evidenced with $\rm \rho \propto -T^{1/4}$ dependence. 

\begin{figure}[t]
	%\centering
	\includegraphics[width=0.45\textwidth]{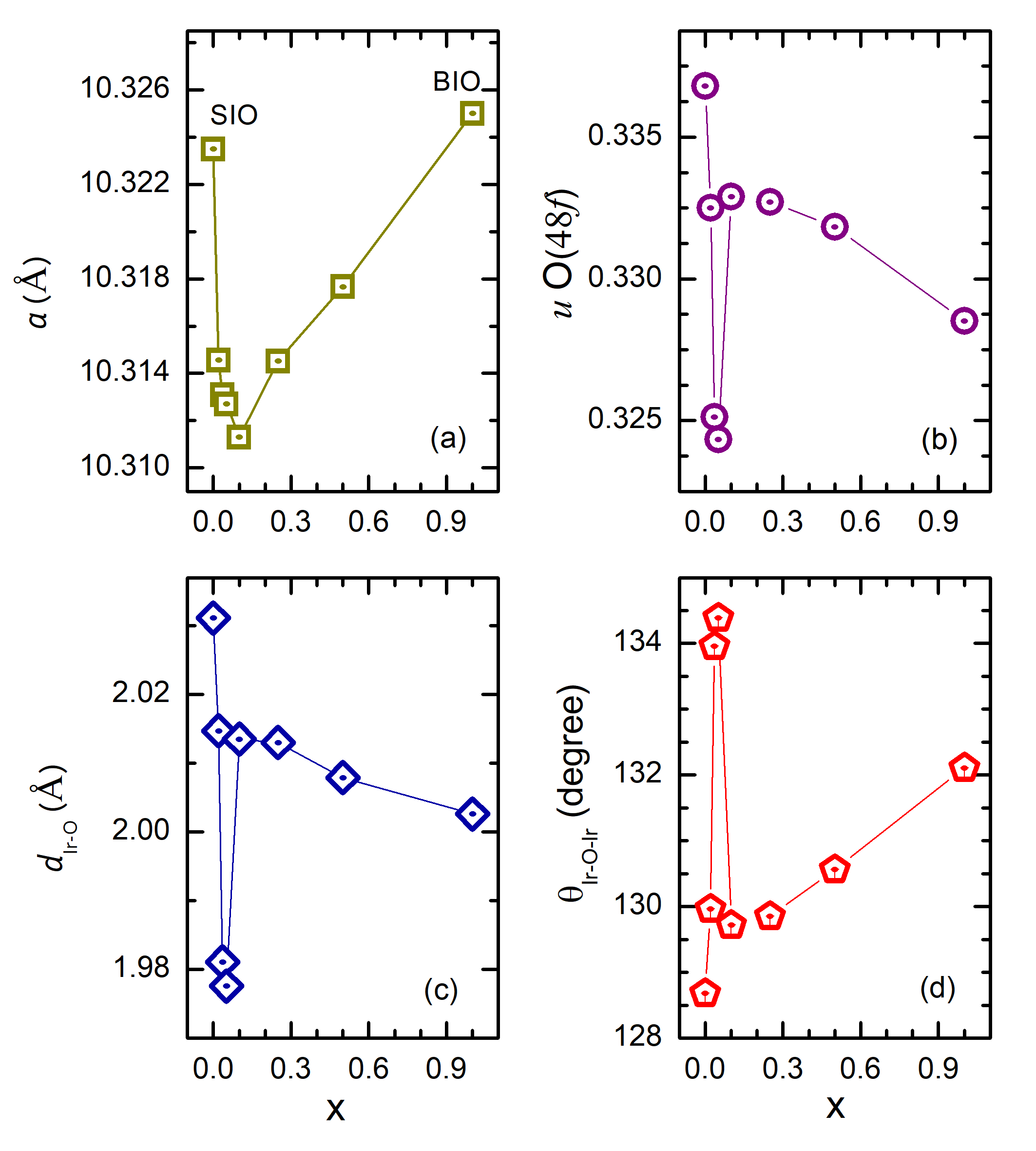}
          \caption {The variation of (a) lattice parameter, (b) O($48f$)-positional parameter ($u$), (c) Ir--O bond distance, and (d) Ir--O--Ir bond angle as a function of $\rm x$ in $\rm (Sm_{1-x}Bi_x)_2Ir_2O_7$. The lines are guide to the eye. The error bars are smaller than the symbol size.}
	\label{latticepara}
\end{figure} 

\textit{Experimental methods} - The samples $\rm (Sm_{1-x}Bi_x)_2Ir_2O_7$ where $\rm x = 0, 0.02, 0.035, 0.05, 0.1, 0.25, 0.5,$ and $1$ were synthesized in air using the method detailed in Ref.~\cite{Telang}. The synchrotron powder X-ray diffraction experiments were performed at 11--BM beamline at the Argonne National Laboratory at a fixed photon energy of 30 keV. The 11--BM beamline couples an efficient Sagittal X-ray beam with a high-precision diffractometer circle and a Si (111) crystal analyser to achieve high sensitivity and resolution. Instrumental resolution at high-Q is better than $\rm \Delta Q/Q \approx 2 \times 10^{-4}$, with a typical $\rm 2\theta$ resolution better than 0.010$^\circ$ at 30 keV. The transmission measurements were performed with the rotating capillary stage to eliminate preferred orientations. The lattice parameters were obtained by Rietveld refinement using the Fullprof suite. The electrical resistivity, magnetic susceptibility and specific heat were measured on dense specimens using the physical property measurements system (PPMS), Quantum Design, USA.\par   

\textit{Results} - The cubic pyrochlore A$ _2 $Ir$ _2 $O$ _7 $ structure consists of four distinct crystallographic sites with A($16d$), Ir($16c$), O($48f$), and O’ ($8b$). The structure has only two-independent variables: the lattice constant $\rm a$, and the x--coordinate ($u$)~of O($48f$). The A--site ion is eight-fold coordinated by six O($48f$) and two O’($8b$) ions. The Ir site is coordinated by O($48f$) alone, forming an octahedron which is trigonally distorted unless $u =$~0.3125.\par

Figure \ref{latticepara}(a) shows the evolution of lattice parameter ($\rm a$) across the (Sm, Bi) series. The values of $\rm a_{SIO}=10.3235~\AA$ for $\rm Sm_2Ir_2O_7$, and $\rm a_{BIO}=10.3250~\AA$ for $\rm Bi_2Ir_2O_7$~are in excellent agreement with literature~\cite{Giampoli, Ueda}. The important point to note here is that the lattice parameter, $\rm a_x$, for all intermediate members is smaller than the end members, attaining the smallest value at $\rm x=0.1$ with $\rm \Delta a = a_x - a_{SIO} \approx -0.012 \AA$. In terms of deviation $\rm \delta$ defined as: $\rm \delta = [(a_x - a_{SIO})/(a_{BIO} - a_{SIO})] \times 100$ this is $\rm \approx -666\%$. As a comparison for analogous (Eu, Bi) series \cite{Telang2019} while $\rm \Delta a$ is of the same order, $\rm \delta$~is only~$\approx -20 \%$. Here, in spite of $\rm a_{SIO} \approx a_{BIO}$, the NLE is highly pronounced. \par  

Figure \ref{latticepara}(b) shows the variation of $u$. An inverse correlation is generally seen between $u$ and  $\rm a$~\cite{Kennedy03}, which is the case for $\rm A_2Ir_2O_7$ where $u =$ 0.339 for Eu$ _2 $Ir$ _2 $O$ _7 $ ($\rm a = 10.2990~\AA$), which decreases to 0.330 for Pr$ _2 $Ir$ _2 $O$ _7 $ ($\rm a = 10.4105~\AA$)~\cite{Millican}. Application of external pressure renders a similar behavior, with increasing pressure leading to an increase in $u$~\cite{Wang2020}. However, in the $\rm (Sm_{1-x}Bi_x)_2Ir_2O_7$ series, both parameters show a decreasing behavior in the NLE region; however, the normal behavior restores for $\rm x > 0.1$. The variations of Ir--O bond length and Ir--O--Ir bond angle are shown in Figure \ref{latticepara}(c) \& \ref{latticepara}(d), respectively. The average Ir--O bond length ($\rm \sim 2~\AA$) is in good agreement with that for IrO$_2$, and is also typical of Ir$^{4+}$ in an octahedral coordination. The Ir--O--Ir bond angle ($\rm \theta_{Ir-O-Ir}$) increases in the anomalous region which is opposite to the trend usually seen, i.e., $\rm \theta$ generally shows a positive correlation with $\rm a$~\cite{Kennedy02}, as is indeed the case for  $\rm x \geq 0.1$.\par   

\begin{figure}[t]
	%\centering
	    \includegraphics[width=0.45\textwidth]{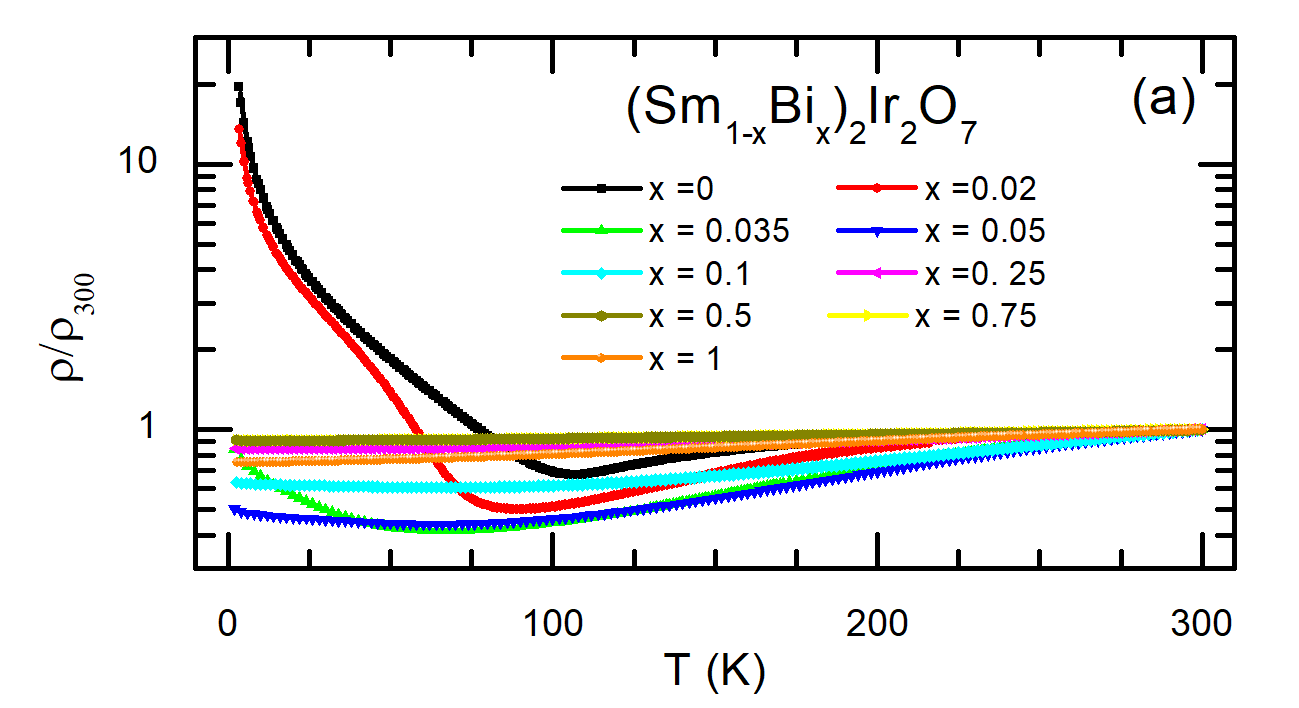}
		\includegraphics[width=0.45\textwidth]{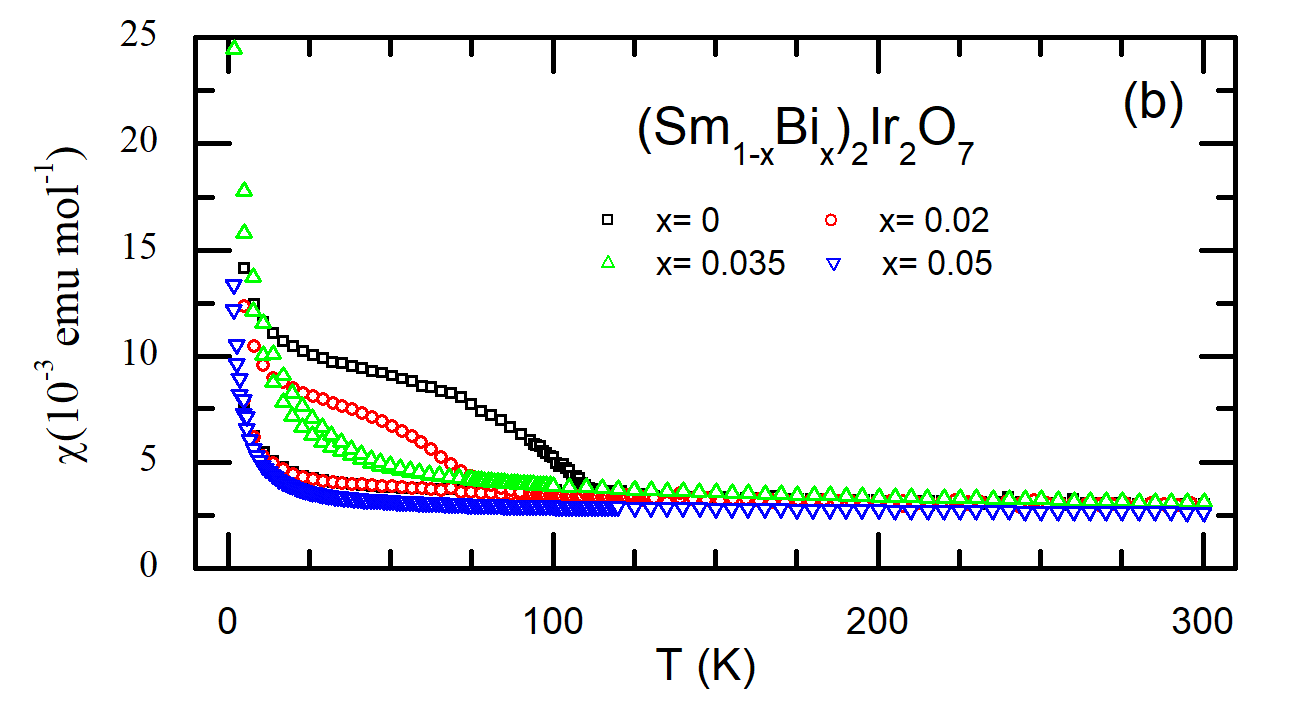}
		\includegraphics[width=0.45\textwidth]{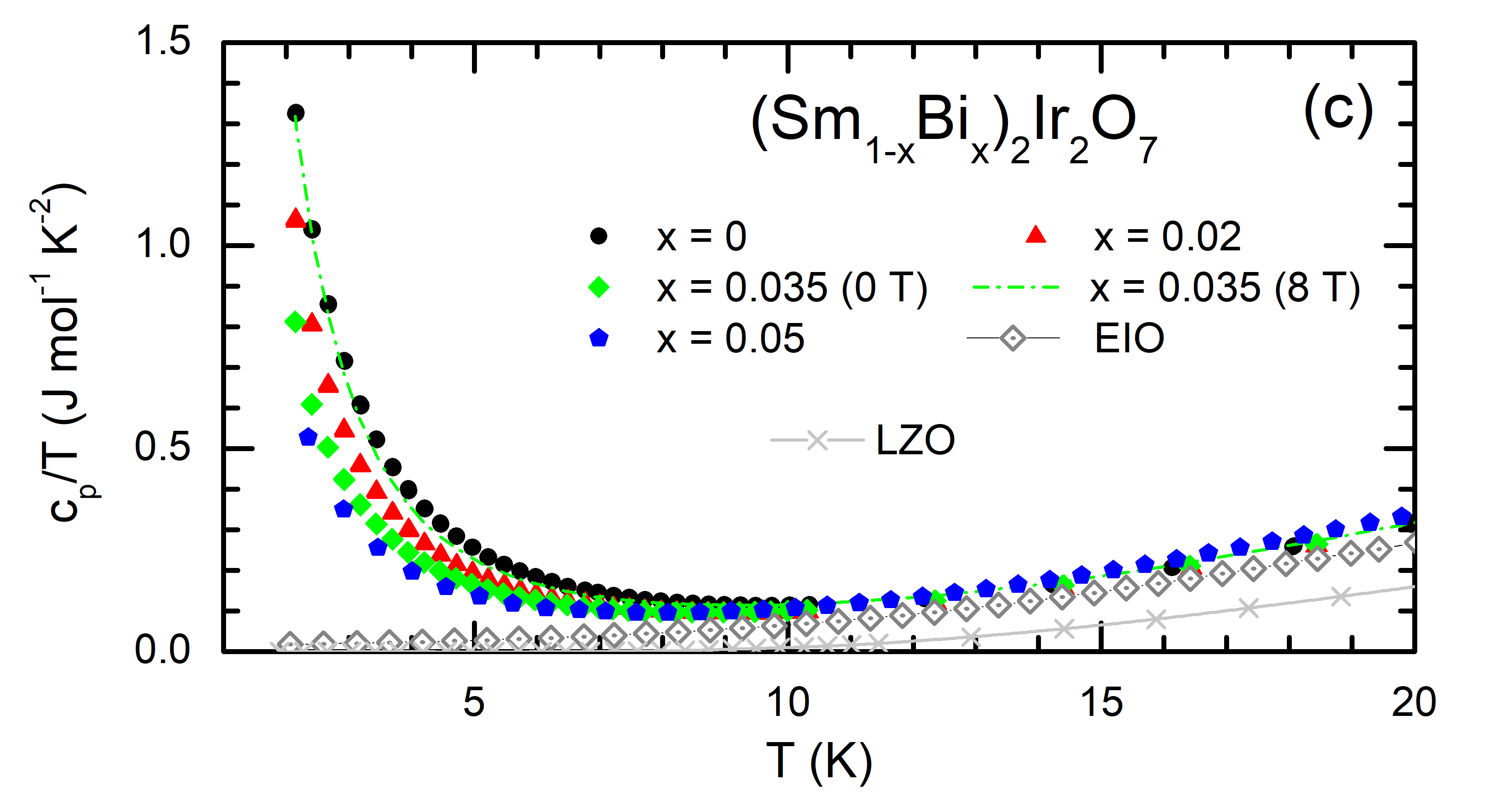}
          \caption {The temperature variation of: (a) normalized electrical resistivity $\rho/\rho_{300}$, (b) magnetic susceptibility ($\chi$), and (c) specific heat divided by temperaure ($\rm c_p/T$). In (c) $\rm c_p/T$ for $\rm Eu_2Ir_2O_7$ (EIO) where $\rm Eu^{3+}$ is non-magnetic, and non-magnetic $\rm La_2Zr_2O_7$ (LZO) are also shown as references. Data for EIO and LZO are adapted from Refs. \cite{Telang2019} and \cite{Singh2008}.}
	\label{Fig2}
\end{figure} 

\begin{figure}[!]
	%\centering
	\includegraphics[width=0.5\textwidth]{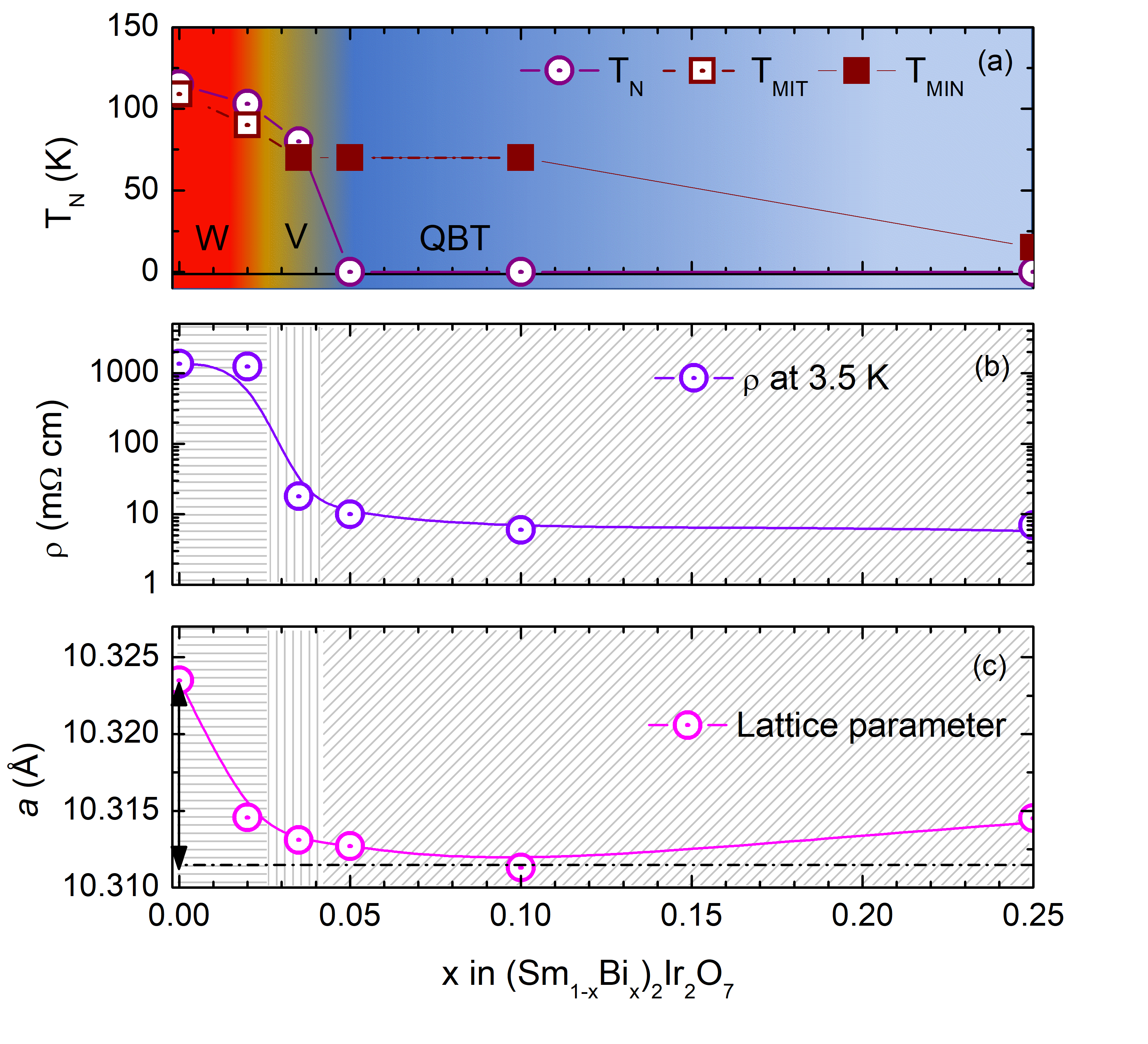}
	\caption {The $\rm x$  dependence of: (a) antiferromagnetic ordering temperature $\rm T_N$, Metal-to-insulator transition temperature ($\rm T_{MIT}$), and the temperature where $\rho$ shows a gradual upturn ($\rm T_{MIN}$); (b) $\rm \rho$ at $\rm T = 3.5~K$; and (c) lattice parameter in $\rm (Sm_{1-x}Bi_x)_2Ir_2O_7$. Here, W represents WSM phase (red or horizontally hatched region) where at low T, $\rm \rho \propto 1/T$; QBT represents quadratic band touching (blue or slanted hatched region) with $\rm \rho \propto -ln T$; V is the new phase (yellow or vertically hatched region) with $\rm \rho \propto -T^\alpha$ ($\rm \alpha \approx 1/4$). Red, yellow and blue colors corresponds to regions 1, 2 and 3 in the main text.}
	\label{Fig3}
\end{figure} 

In Fig. \ref{Fig2}, the  normalized electrical resistivity $\rm \rho / \rho(300)$, magnetic susceptibility $\chi$, and specific heat $\rm c_p$ are shown. The pristine sample shows a sharp MIT concomitant with the AIAO ordering. The value of $\rm T_N = 115~K$ is taken as the point of ZFC--FC bifurcation in $\chi$. With increasing $\rm x$ both $\rm T_{MIT}$ and $\rm T_N$ are suppressed,  and $\rho$ decreases in magnitude.The magnetic behavior in the paramagnetic state is weakly temperature dependent. No signs of Sm moments ordering could be seen in our bulk measurements in agreement with previous studies~\cite{Graf2014, Donnerer}. The effective moment on Sm$^{3+}$ ($\rm 0.2~\mu_B$) is found to be small and the interaction between Sm moments weak ($\rm \theta_p < 1~K$), in agreement with literature~\cite{Graf2014}. As shown in Fig. \ref{Fig2}(c), $\rm c_p/T$ curves show an upturn below 10~K, which is due to the short-range ordering of Sm-moments analogous to other Sm pyrochlores~\cite{Singh2008, Mauws2018}. With increasing $\rm x$, the upturn in $\rm c_p/T$ is suppressed to lower temperatures. $\rm c_p/T$ of $\rm Eu_2Ir_2O_7$, where Eu$^{3+}$ is nonmagnetic, is also shown to illustrate that the low-temperature $\rm c_p$ is dominated by the Sm sublattice and that the infarction between the two sublattices is weak (see Supplementary). This is in contrast with $\rm Sm_2Mo_2O_7$ where the Mo orders in a two-in/two-out magnetic state inducing a two-in/two-out state on the Sm sublattice at a relatively high temperature of $\approx$15 K~\cite{Singh2008A}. 

\begin{figure}[!]
	\centering
	\includegraphics[width=0.48\textwidth]{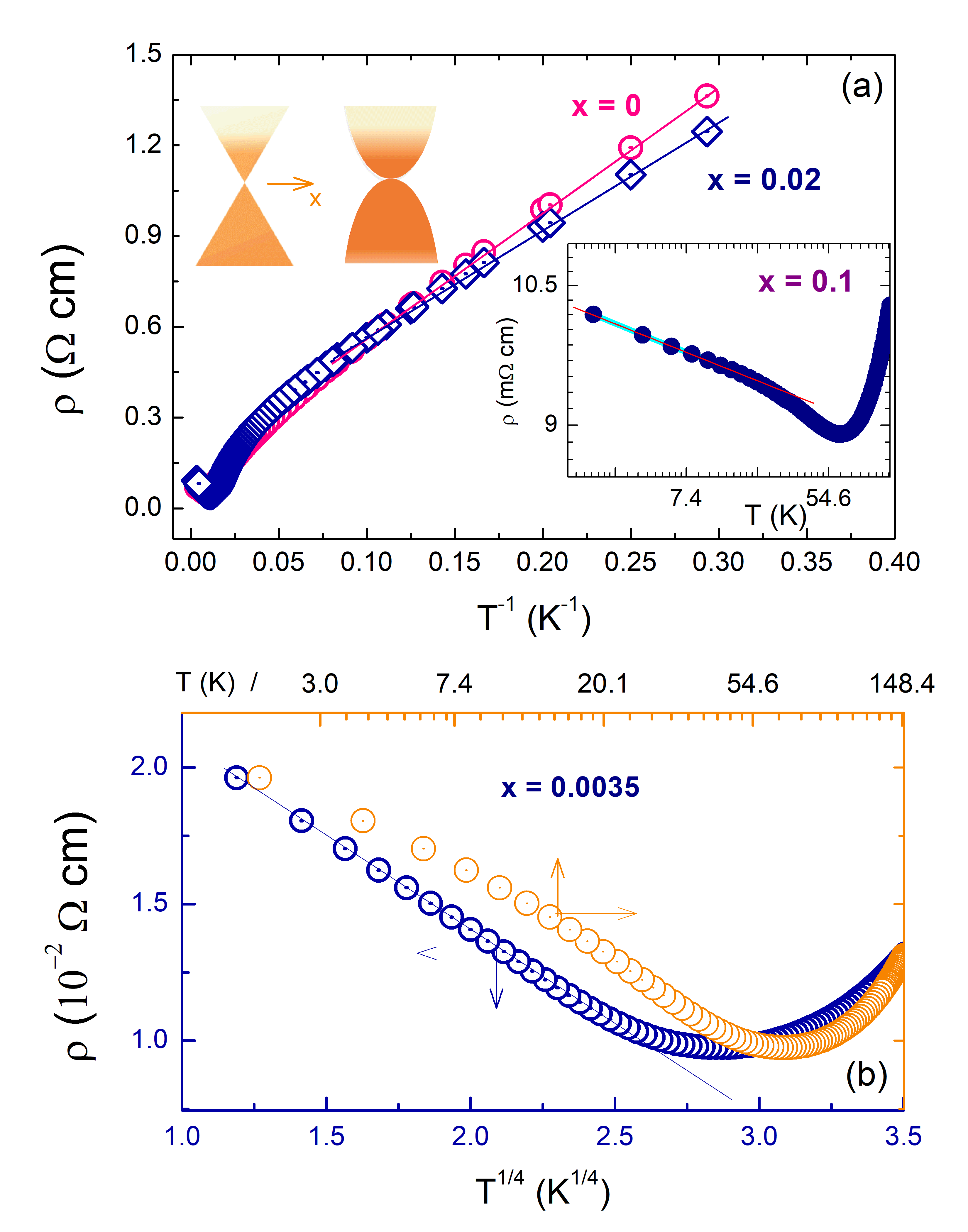}
	\caption {(a) $\rm \rho$ is shown as function of $\rm 1/T$ for $\rm x = 0$, and $\rm 0.02$. The lines depict a $\rm 1/T$ dependence. Lower inset: $\rm -ln T$ dependence at $\rm x = 0.1$. Upper inset: schematic band structure with and without Bi doping. The linearly dispersing Weyl nodes near $\rm E_F$ at $\rm x = 0$ (left) are replaced by QBT at $\rm x \geq 0.02$ (right). Note that the Weyl nodes appear in pairs with right/left-hand chiralities; here, for simplicity, only a single node is shown. (b) The temperature variation of $\rho$ at $\rm x = 0.035$ plotted as a function of $\rm T^{1/4}$ (bottom axis) and $\rm ln T$ (top axis). The x-axis and y-axis range for both scalings is kept identical. The lines are drawn as a guide to the eye.} 
	\label{Fig4}
\end{figure}

The evolution of $\rm T_N$, $\rm T_{MIT}$, and $\rm \rho (3.5 K)$ with $\rm x$ is shown in \ref{Fig3}. We broadly classify the electronic properties into three regions: 
The region (1) extends up to $\rm x \approx 0.02$. In this region, the lattice parameter decreases most dramatically (Fig. \ref{Fig3}c). In fact, 90\% of the total observed decrease happens in this region. However, it is  intriguing that despite a significant NLE $\rm T_N$, $\rm T_{MIT}$, and $\rho(3.5~K)$ remain relatively unchanged. The low-temperature $\rho$ in this region varies as {\rm $1/T$} (Fig.~\ref{Fig4}a). 

The region (3) extends from $\rm x \approx 0.05$ to $x \approx 0.25$.  Here, $\rm a$ varies relatively slowly but  $\rho(3.5~K)$ drops by more than two orders of magnitude compared to region (1). Simultaneously, the sharp MIT is replaced by a gradual upturn below $\rm T_{MIN}$, which remains non-zero up to $\rm x \approx 0.25$; however, and notably, $\rm T_N$ drops to 'zero' in this region (Fig. \ref{Fig3}a). Here, the low-temperature upturn varies as $\rm -ln T$ as shown in Fig. \ref{Fig4}a (inset) for $\rm x = 0.1$. Fitting for the other samples is shown in the supplementary.    

The region (2) is a narrow region sandwiched between regions (1) and (3). In this region, the low--temperature $\rm \rho$ follows a peculiar $\rm -T^\alpha$ dependence with $\rm \alpha \approx 1/4$ (see Fig.~\ref{Fig4}b). Attempts to fit the data using $\rm -lnT$ or $\rm 1/T$ or using the Arrhenius or VRH (with or without correlations) models did not yield satisfactory fit, suggesting that it is a 'new' phase. The decrease in T$_N$ is most pronounced in this region; in fact in the narrow range $\rm 0.035 \leq x \leq 0.05$, $\rm T_N$ drops from near 70 K to below our lowest measurement temperature ('zero'). In the (Eu, Bi) series, no new phase could be identified as the boundary region there is relatively broad.   

\textit{Discussion} - Typically, in a solid solution the lattice parameter for intermediate compositions follow a linear behavior, which is referred to as the Vegard’s law~\cite{Vegard}. While this law may not always be obeyed perfectly, it guides the evolution of lattice constants in a solid solution. Minor deviations from the Vegard’s law may arise both for metal alloys~\cite{Murphy, Axon,Lubarda} and metal oxides~\cite{Ganguly, West, Baidya}. However, such deviations have been mainly in the form of either change of slope~\cite{Nyon,Baidya}, departure from a linear behavior~\cite{Kong, Dismukes, West}, and at times only along a certain crystallographic axis~\cite{Ganguly, Hogan2015}. Further, the magnitude of these deviations is generally small and are typically ascribed to such effects as clustering, phase segregation, limited solid solubility, and valence fluctuation. 

Our synchrotron data does not shown any evidence of clustering or phase segregation (see supplementary material). While it is not uncommon among the compounds of Sm to have a mixed valence state, presence of Sm$^{2+}$ for Sm$^{3+}$ would only result in an increase in $\rm a$ as Sm$^{2+}$ ($\rm 1.27~\AA$) is larger than Sm$^{3+}$ ($\rm 1.08~\AA$)~\cite{Shannon}. In fact, in an analogous (Eu, Bi) series, no evidence for any change in the valence of either Eu, Bi or Ir could be found for any x~\cite{Telang2021}. A +3 oxidation state for Bi has also been shown in $\rm Bi_2Ir_2O_7$ using XANES/XPS~\cite{Sardar}. Since all our samples were prepared under identical condition, variable oxygen off-stoichiometry can also be ruled out. Similarly, the stereochemical activity of 6s$^2$ lone-pair, if present, should be more pronounced in the Bi--rich samples, however, the NLE has been limited to dilute Bi doping only. Thus, the NLE appears to be an electronic effect which is strong enough to overwhelm the steric effect for small $\rm x$.

That the electronic states play a role in deciding the lattice parameter in these pyrochlores is evident from the fact that $\rm a_{SIO}=10.3235~\AA$ is nearly as large as $\rm a_{BIO}=10.3250~\AA$ even though Sm$^{3+}$ (1.079~\AA) is significantly smaller than Bi$^{3+}$ (1.170~\AA). Intriguingly, the lattice parameter decrease still further, and by a significant amount, when as small as 2 \% of Bi is doped in $\rm Sm_2Ir_2O_7$. Such a behavior is not observed in the pyrochlore stannates where the Vegard’s law follows nicely \cite{Telang2019, Kennedy03}. 

To gain a qualitative understanding of this unusual behavior we recall previous theoretical calculations by Krempa et al. \cite{Krempa2012, Krempa2013}. They have shown that besides the Ir--Ir hopping mediated via O($48f$), the direct Ir--Ir hopping which can be relatively large due to an extended nature of Ir’s 5d orbitals plays a crucial role in realizing various topological phases in the pyrochlore iridtates. In particular, for a given Hubbard U, which is fixed by the choice of $\rm A$ in $\rm A_2Ir_2O_7$, the WSM phase has been shown to stabilize over a narrow range of values of the transfer integrals ratio $\rm t_\sigma / t_{oxy}$, where $\rm t_\sigma$ and $\rm t_{oxy}$ are the hopping integrals associated with direct Ir--Ir hopping and Ir--Ir hopping via O($48f$) respectively. 

We now consider the following scenario: Had the lattice expanded normally upon small Bi doping, both Ir--Ir bond distance ($\rm d_{Ir-Ir} =  (\sqrt{2} a)⁄4$ ) and Ir--O--Ir bond angle would have shown an increasing trend~\cite{Kennedy02}, as is the case for $\rm x > 0.1$ in Fig.~\ref{latticepara}. But increasing Ir--Ir distance decreases $\rm t_\sigma$, and at the same time increasing Ir--O--Ir bond angle increases $\rm t_{oxy}$, i.e, the ratio $\rm t_\sigma / \rm t_{oxy}$ will change more substantially in this case than when both either increase or decrease. Indeed, in the region of NLE while Ir--Ir decreases and Ir--O--Ir increases, both $\rm t_\sigma$ and $\rm t_{oxy}$ increase together, preserving the Weyl phase. This is evident from Fig. \ref{Fig4}a where we show $\rho$ for $\rm x = 0$ and $\rm 0.02$ samples as a function of $\rm 1/T$. As can be seen both samples exhibit a $\rm 1/T$ behavior at low temperatures, which is one of the hallmarks of the Weyl phase~\cite{Hosur}. In the Weyl phase, the electron-hole symmetry about the Weyl node results in current carrying states with zero total momentum leading to this characteristic $\rm 1/T$ dependence. A minor change in U due to 2\% doping can ignored; however, in the region where lattice expands, both $t_\sigma$ and U will decrease pushing the system into the metallic region of the phase diagram shown in Fig. 2 in Ref. \cite{Krempa2012}, as is also found to be the case here experimentally.  

In region (3), while $\rm T_N$ drops to zero, the resistivity upturn remains (Fig. \ref{Fig3}a). Here, low temperature $\rho$ follows a $\rm -ln(T)$ dependence, signature of a nodal non-Fermi liquid behavior associated with QBT~\cite{Lucile2014, Kondo2015}. Savary et al.~\cite{Lucile2014} had theoretically argued on the possibility of tuning the ground state continuously from WSM to QBT, but with a quantum critical point (QCP) located in between. With Bi doping we indeed see a new phase in the boundary region (labeled 'V' in Fig. \ref{Fig3}a). In this phase, $\rho$ neither varies as $\rm 1/T$ (Weyl) nor as $\rm -ln T$ (QBT) (other scenarios are also ruled out, see Supplementary), but it rather follows a $\rm -T^{1/4}$ dependence over a broad temperature range, which could be a finite temperature manifestation of the QCP expected from theory. Under external pressure a complete suppression of AIAO takes place under 6 GPa~\cite{Wang2020}; however, with Bi doping not only the chemical pressure due to NLE, but Ir (5d) - Bi (6s/6p) hybridization also contributes to the electronic states near the Fermi energy leading to a richer phenomenology~\cite{Qi_2012}.   

In conclusion, we have shown the occurrence of a NLE when as low as 2 \% of Sm$^{3+}$ ions in Sm$ _2 $Ir$ _2 $O$ _7$ are substituted by larger Bi$^{3+}$ ions, protracting the Weyl phase into the Bi doped region. At higher doping, the WSM phase collapses and the phase with QBT emerges. At the interface of these two phases, distinct signature of a new phase with a peculiar $-\rm T^{1/4}$ dependence of $\rho$ is shown to persist over a broad temperature range indicating proximity to the QCP theoretically expected~\cite{Lucile2014}. Both these intriguing results warrant further experimental and theoretical studies.    

\section*{Acknowledgments}
We would like to thank Prasenjit Ghosh for useful discussion, and Abhisek Bandyopadhyay for help in analyzing the EXAFS data in the analogous (Eu, Bi) series which confirmed that a valence change or mixed valent state is not present. SS would like to that SERB for funding under the core-research grant EMR/2016/003792/PHY. PT and SS would like to thank the rapid access beamtime facility at Argonne National laboratory for the synchrotron X-ray diffraction experiments.

\nocite{}
\section*{References}   
\begin{figure*}[!]
	%\centering
	\includegraphics[width=1\textwidth]{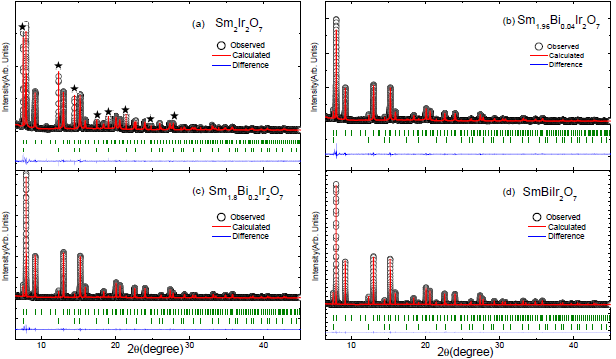}
	%\captionsetup{justification=raggedright,singlelinecheck=false}
	\caption{Representative Rietveld refinement plots for the Sm$_{2-2x}$Bi$_{2x}$Ir$_2$O$_7$ series. A mixed phase refinement was carried out consisting of: (I) Pyrochlore phase, and (II) high-purity Si, which was added as an internal standard.}
	\label{RR}
\end{figure*}

\begin{figure*}[!]
	%\centering
	\includegraphics[width=0.50\textwidth]{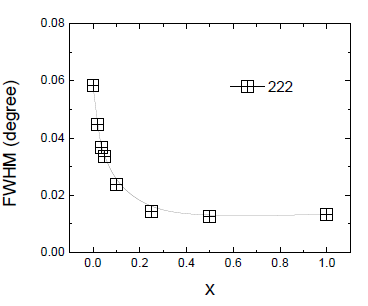}
	%\captionsetup{justification=raggedright,singlelinecheck=false}
	\caption{Variation of FWHM of the highest-intensity diffraction peak 222. The FWHM decreases monotonously with increasing x. Note that the effect of strain has not been subtracted. The line through the data points is a guide to the eye. The error bars are less than the size of the symbol. }
	\label{FWHM}
\end{figure*}

\bibliography{References}
\bibliographystyle{revtex}
\pagebreak
\pagebreak
\newpage
\textbf{Rietveld refinement}\\
The Rietveld refinement results are tabulated in Table. I. In Fig. \ref{RR}, the Rietveld refinement plots are shown for a few representative samples. Two phase refinement was done with Si as second phase which was added as an internal standard for greater accountability of the lattice parameters. The FWHM for the 222 peak is shown in Fig. \ref{FWHM} as representative. With Bi doping FWHM decreases monotonically all through the doping range. The lattice parameter and various bond distances are given Table II (see last page).\\ 

\textbf{Resistivity}\\
We carefully analyzed the resistivity of each sample using six different scaling behaviors: (i) 1/T, (ii) T$^{1/4}$, (iii) $\rm -lnT$, (iv) Variable Range Hopping (VRH), (v) VRH in the presence of correlations, and (vi) Arrhenius. Here, we show the results for first four. The fittings (iv) and (v) gave mare or less similar results (in terms of the quality of fit), and (vi) could not fit the data at all. As explained through the figure captions, in region (1) ($\rm 0 \leq x \leq 0.02$), we find 1/T dependence to be clearly superior over all the other scalings. In region (2) x = 0.035, a clear $\rm -T^{1/4}$ dependence is seen (no other model could fit the resistivity data as good as $\rm -T^{1/4}$). In region (3), the $\rm -lnT$ dependence is seen in the beginning (up to x = 0.1) but as x increases the VRH type behavior begins to emerge. \\

\textbf{Magnetic susceptibility}\\
In $\rm Sm_2Ir_2O_7$, the magnetic transition around 117 K is due to the AIAO ordering of the Ir moments. Besides this transition no other magnetic transition has been found in the bulk measurements either in this or the previous studies. However, using $\mu_{SR}$ some evidence for Sm moments ordering below 10 K was reported by Asih et al. [J. Phys. Soc. Jpn. 86, 024705 (2017)] who estimated the ordered Sm moment to be close to 0.1 $\rm \mu_B$ but this does not agree with other studies. Graf et al. analyzed the low temperature DC susceptibility of $\rm Sm_2Ir_2O_7$ using the modified Curie Weiss law [Journal of Physics: Conference Series 551 (2014) 012020] and reported the effective moment on Sm to be close to 0.2$\rm \mu_B$. We took the same approach and fitted the low-temperature data using the modified Curie-Weiss law: $\rm \chi= \chi_0+  C⁄(T- \theta_p)$, here $\rm \chi_0$ is the temperature independent term, C is the Curie constant and $\rm theta_p$ is the Weiss temperature. Our results are in agreement with Graf et al... The value of $\chi_0$ ranges from 0.003 to 0.002 emu mol$^{-1}$. A similar value is obtained by fitting $\chi$(T) above the ordering temperature $T_N$. The value of $\rm \theta_p < 1 K$ suggests that the effective interaction between Sm moments is weak. The value of C ranges from 0.02-0.03~$\rm emu~mol^{-1}~K^{-1}$.  \\

\begin{figure*}[!]
	%\centering
	\includegraphics[width=0.48\textwidth]{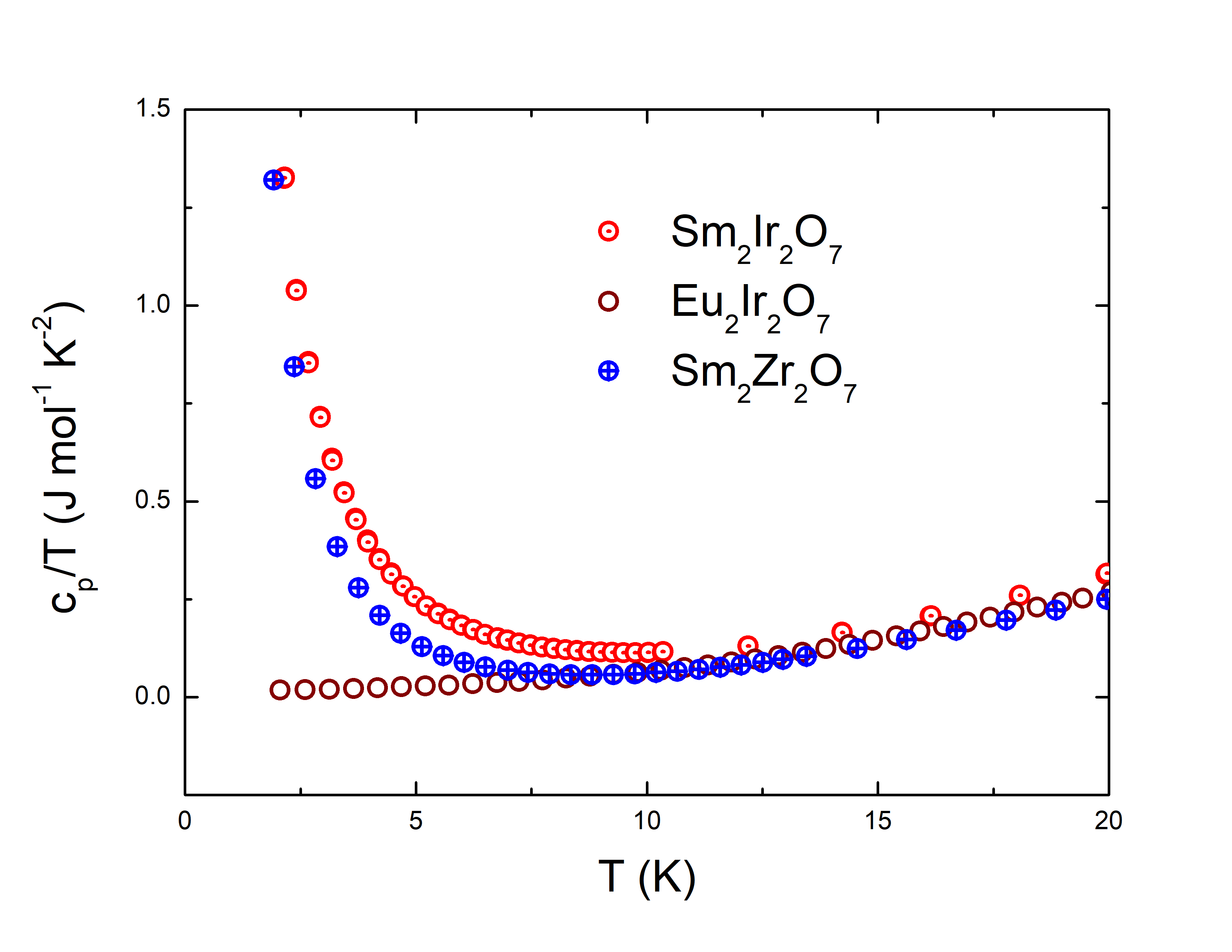}
	\includegraphics[width=0.48\textwidth]{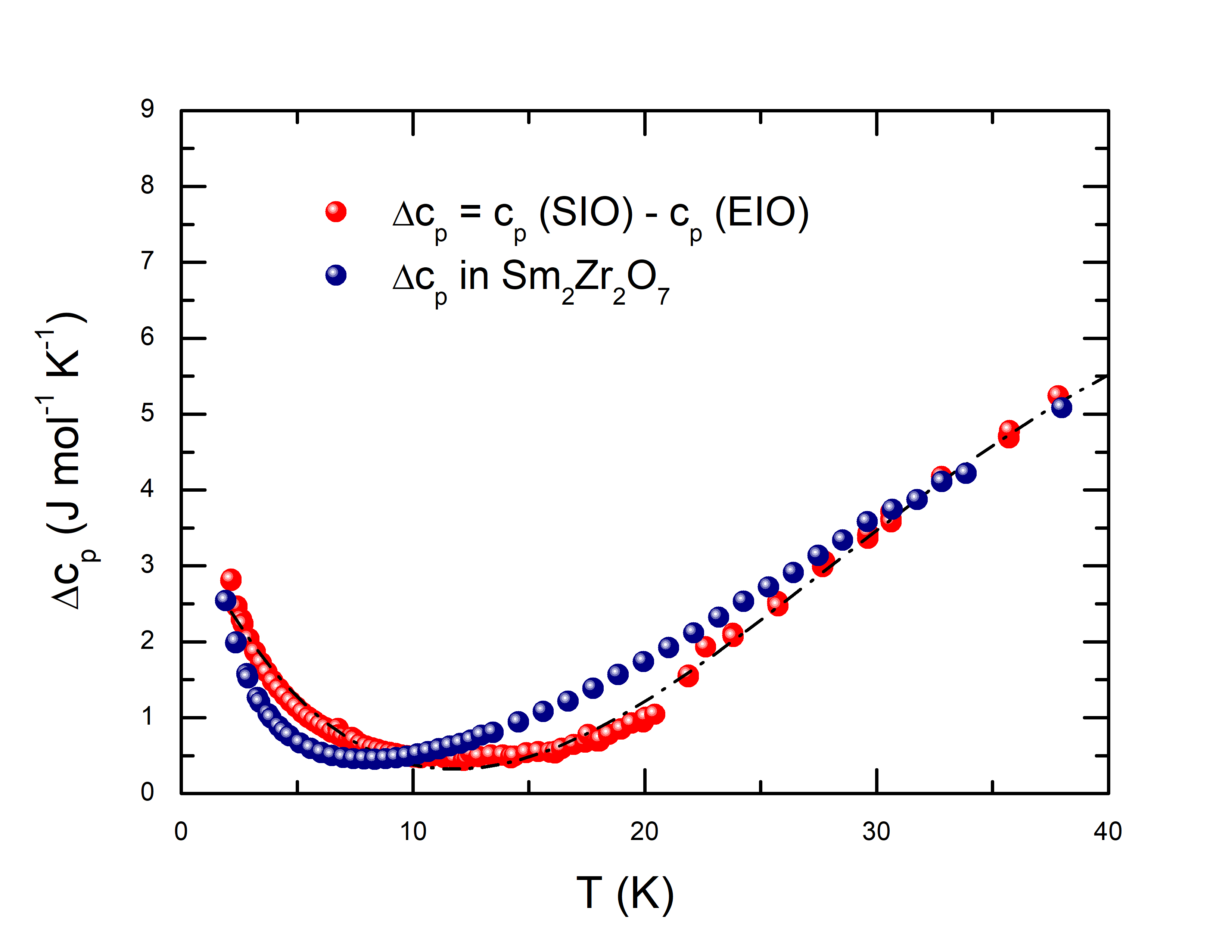}
	%\captionsetup{justification=raggedright,singlelinecheck=false}
	\caption{Temperature variation of specific heat ($\rm c_p$) in $\rm Sm_2Ir_2O_7$, $\rm Eu_2Ir_2O_7$, and $\rm Sm_2Zr_2O_7$ (left), and (b) Specific heat associated with Sm sublattice ($\rm \Delta c_p$). $\rm La_2Zr_2O_7$ is used as a lattice template for estimating $\rm \Delta c_p$ in $\rm Sm_2Zr_2O_7$ adapted from Singh et al. [Phys. Rev. B 77, 054408 (2008)].}
	\label{cpsupp}
\end{figure*}

\textbf{Specific heat}\\
The specific heat ($\rm c_p$) of $\rm Sm_2Ir_2O_7$ below T = 15 K is in good agreement with that of $\rm Sm_2Zr_2O_7$ (see Fig. \ref{cpsupp}). This suggests that in $\rm Sm_2Ir_2O_7$ the Ir ordering has only a small effect on the magnetic behavior of Sm-sublattice. To estimate the contribution of Sm 4$f$ electrons to the measured specific heat, we used specific heat of $\rm Eu_2Ir_2O_7$ to represent phonon contribution and the contribution due to the ordered Ir moments. This admittedly is only a crude way of estimating $\rm \Delta c_p$ but nevertheless gives a good qualitative estimate. The justification for using  $\rm Eu_2Ir_2O_7$ as a template is that the Ir's AIAO ordering temperature is equal within $\pm$5 K for the two systems, and in $\rm Eu_2Ir_2O_7$, the Eu ion does not carry any magnetic moment. In doing so what we have overlooked however is the weak Sm-Ir exchange. The presence of weak Sm-Ir exchange can be seen in the Fig. \ref{cpsupp}, where  $\rm \Delta c_p$  for the Ir sample is less steeper than for the Zr sample.

\newpage

\begin{table*}
	\caption{The goodness-of-fit parameters $\chi^2$, $R_p$, and $R_wp$ for the Rietveld refinement of various samples with Bi composition x shown as \%  (x = 0 corresponds to Sm$_2$Ir$_2$O$_7$ and x = 100 to Bi$_2$Ir$_2$O$_7$). For lattice parameters and bond distance see Table II at the end}
	\label{Table1}
	\begin{center}
		\begin{tabular}{c | c | c | c }
			\hline
			\hline	
			\hspace*{0.2cm} $x$  \hspace*{0.2cm} &  \hspace*{0.3cm} $\chi$$^2$  \hspace*{0.3cm} &   \hspace*{0.3cm} $R_p$  \hspace*{0.3cm} &  \hspace*{0.3cm} $R_{wp}$  \hspace*{0.3cm} \\	
			\hline
			\hline  
			
			$x=0$  & \hspace*{1cm}$ 1.68$ \hspace*{1cm} & $5.92$ & 8.37 \\ 
			\hline 
			%\vspace{1mm}
			$x=2 $ & \hspace*{1cm}$2.35$ \hspace*{1cm} & $6.24$ & 8.82 \\
			\hline 
			%\vspace{1mm}
			%\vspace{1mm}
			\hspace*{0.3cm}$x=3.5$ &  \hspace*{1cm}$5.47$ \hspace*{1cm} & $9.16$ & 15.1 \\ 
			\hline 
			%\vspace{1mm}
			$x=5$ &  \hspace*{1cm}$6.76$ \hspace*{1cm} & $10.1$ & 16.1 \\
			\hline 
			%\vspace{1mm}
			\hspace*{0.2cm}$x=10$ &  \hspace*{1cm}$2.64$ \hspace*{1cm} & $6.57$ & 9.6 \\
			\hline 
			%\vspace{1mm}
			\hspace*{0.2cm}$x=25$ &  \hspace*{1cm}$2.16$ \hspace*{1cm} & $5.5$ & 8.06 \\
			\hline 
			\hspace*{0.2cm}$x=50$ &  \hspace*{1cm}$1.71$ \hspace*{1cm} & $4.87$ & 6.99\\
			\hline 
			%\vspace{1mm}
			%\vspace{1mm}
			\hspace*{0.6cm}$x=100$ \hspace*{0.2cm}&  \hspace*{1cm}$1.98$ \hspace*{1cm} & $4.71$ & 6.18 \\
			\hline \hline 
		\end{tabular}
	\end{center}
\end{table*}

\begin{figure*}[!]
	%\centering
	\includegraphics[width=0.48\textwidth]{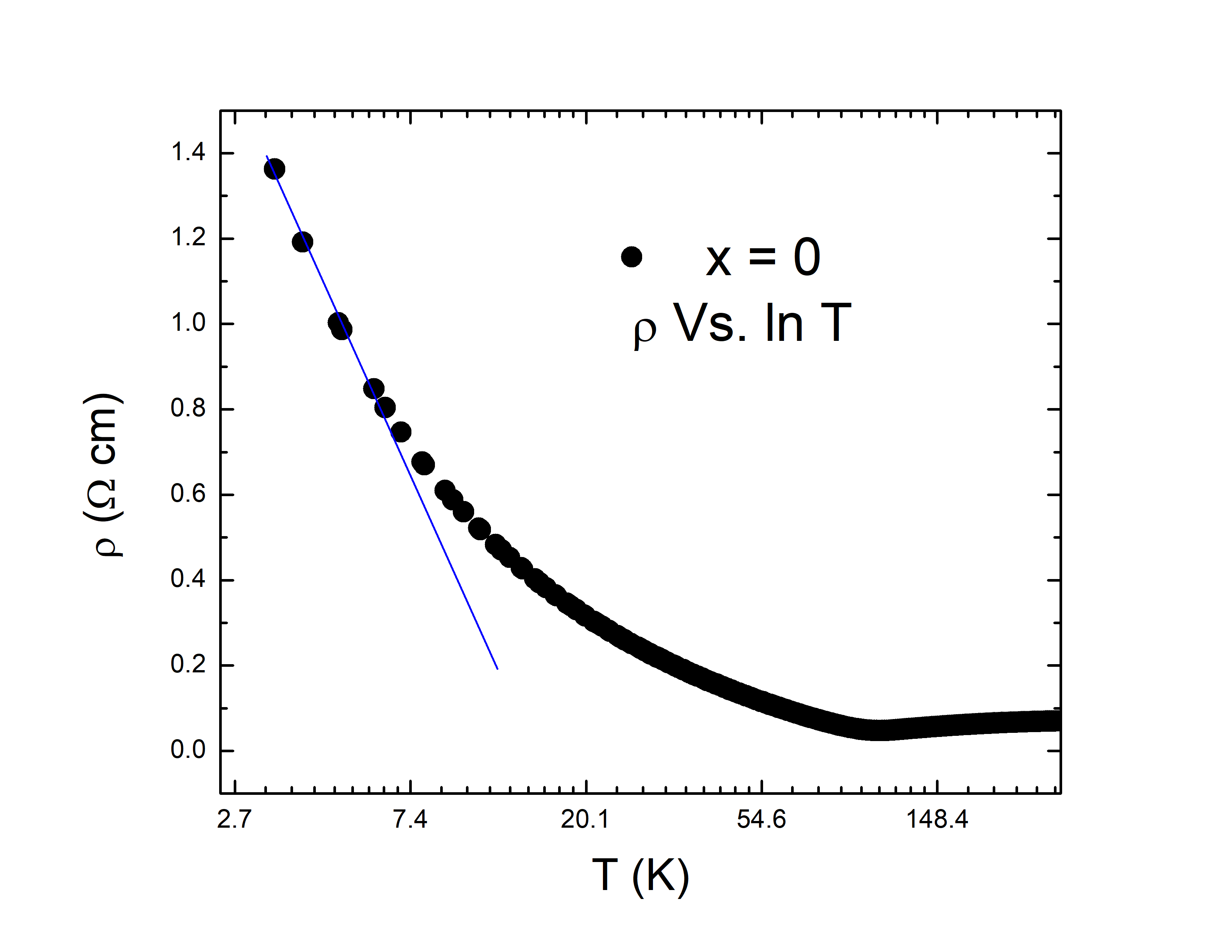}
	\includegraphics[width=0.48\textwidth]{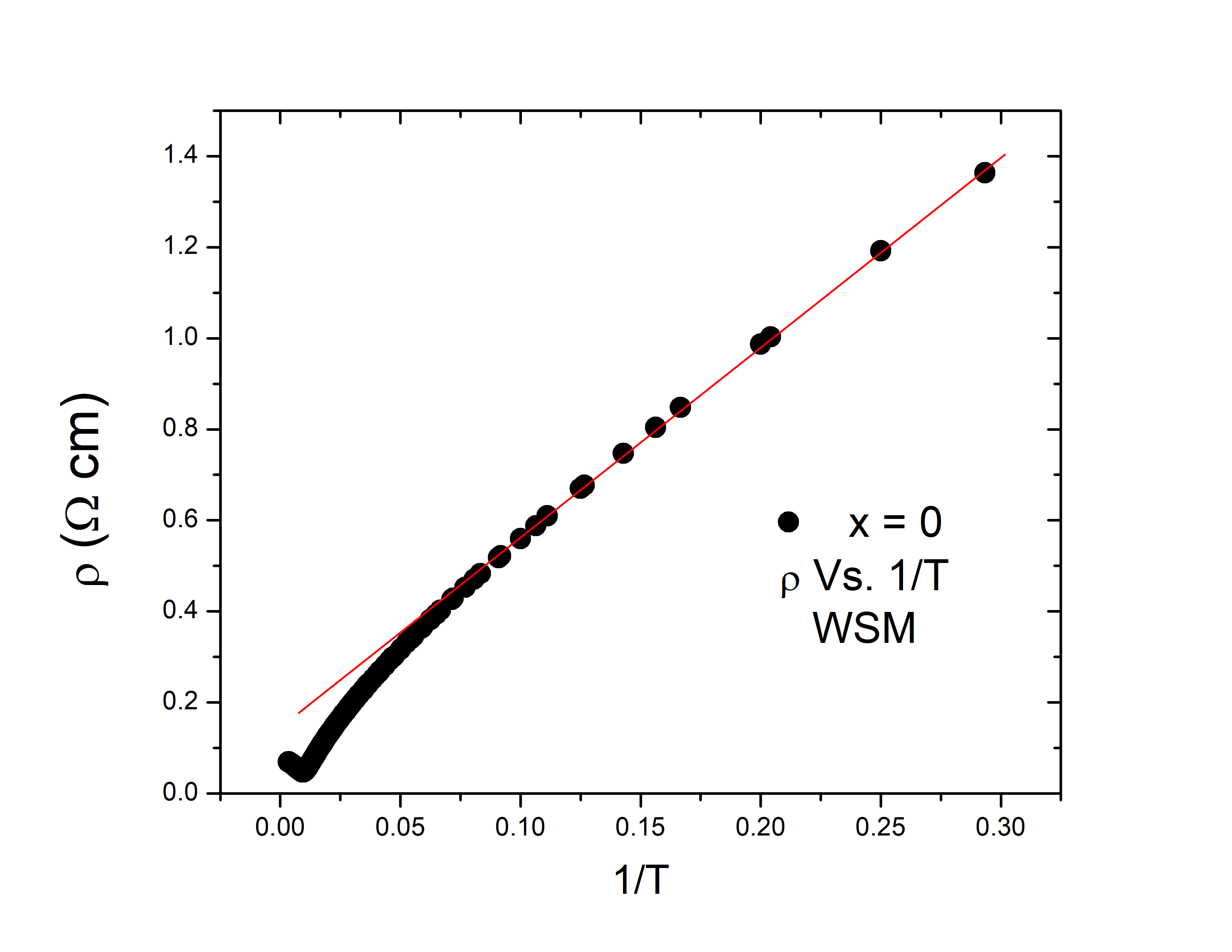}
	\includegraphics[width=0.48\textwidth]{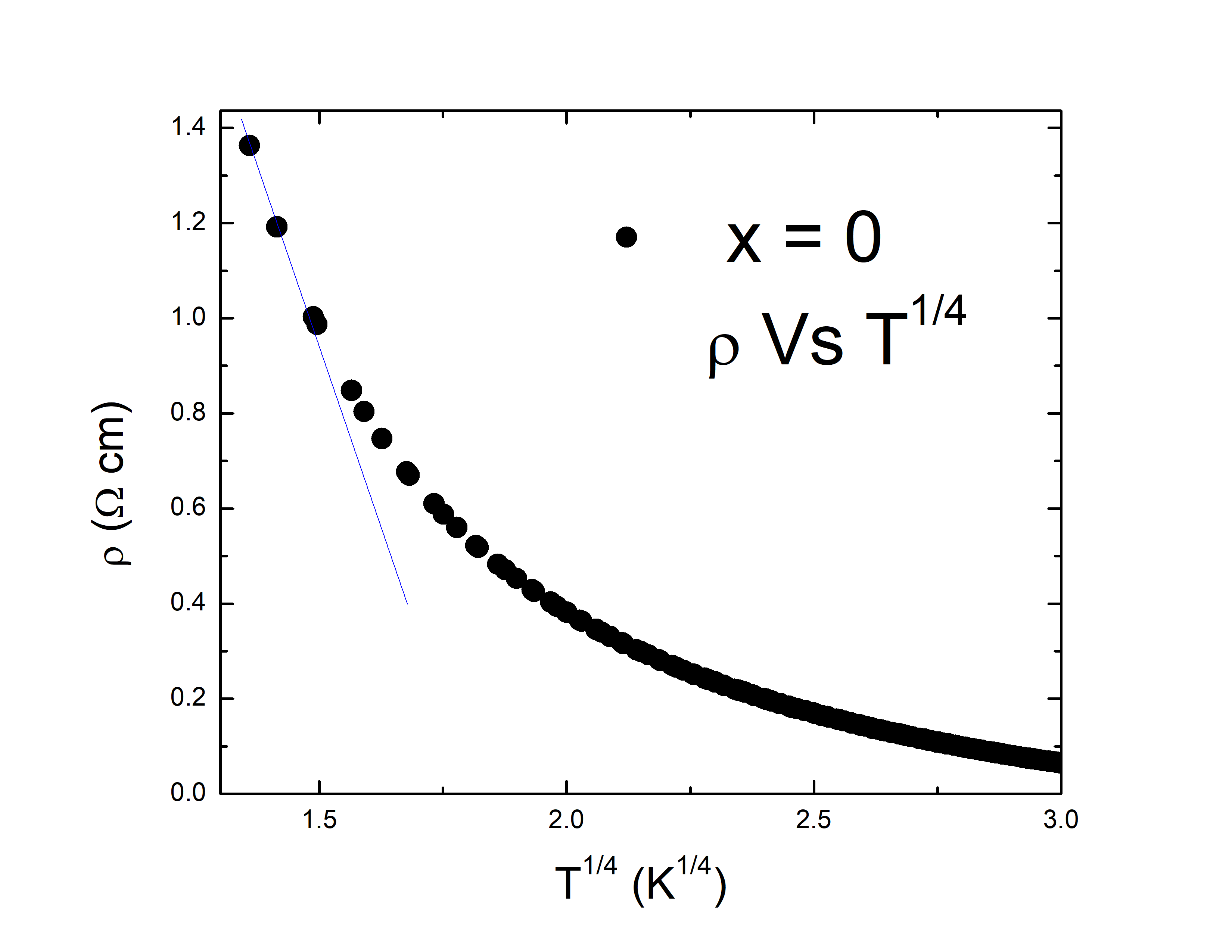}
	\includegraphics[width=0.48\textwidth]{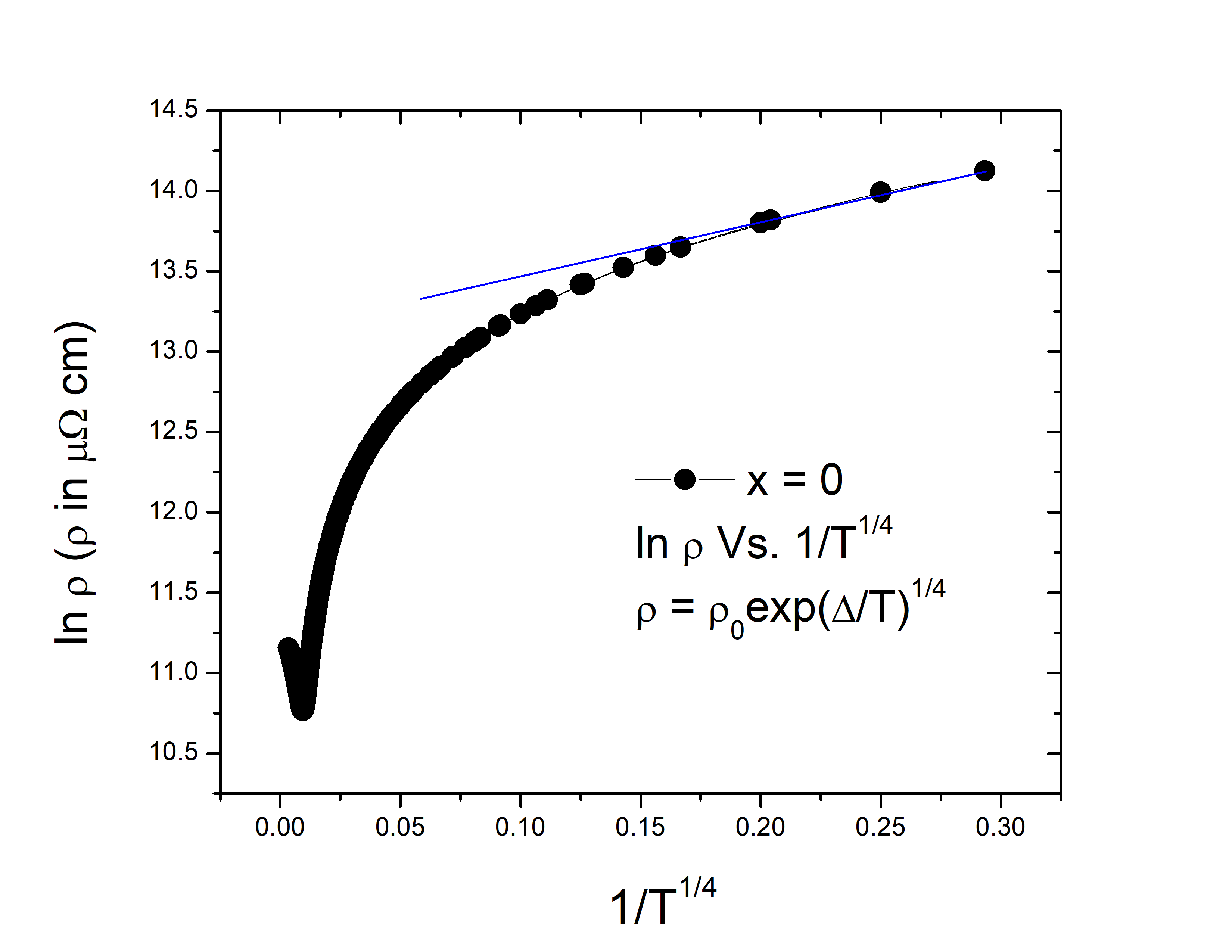}
	%\captionsetup{justification=raggedright,singlelinecheck=false}
	\caption{This figure shows the comparison of various scaling behaviors for x = 0. Evidently, only when plotted as 1/T best linear variation is seen; for all other cases the variation is either non-linear or the range of linearity is very small. Hence, it is reasonable to conclude that the low-temperature resistivity of $\rm Sm_2Ir_2O_7$ has a 1/T dependence expected for a Weyl semimetallic phase as explained in the main text.  Similar graphs were obtained for x = 0.02 (not shown).}
	\label{rho0}
\end{figure*}

\begin{figure*}[!]
	%\centering
	\includegraphics[width=0.48\textwidth]{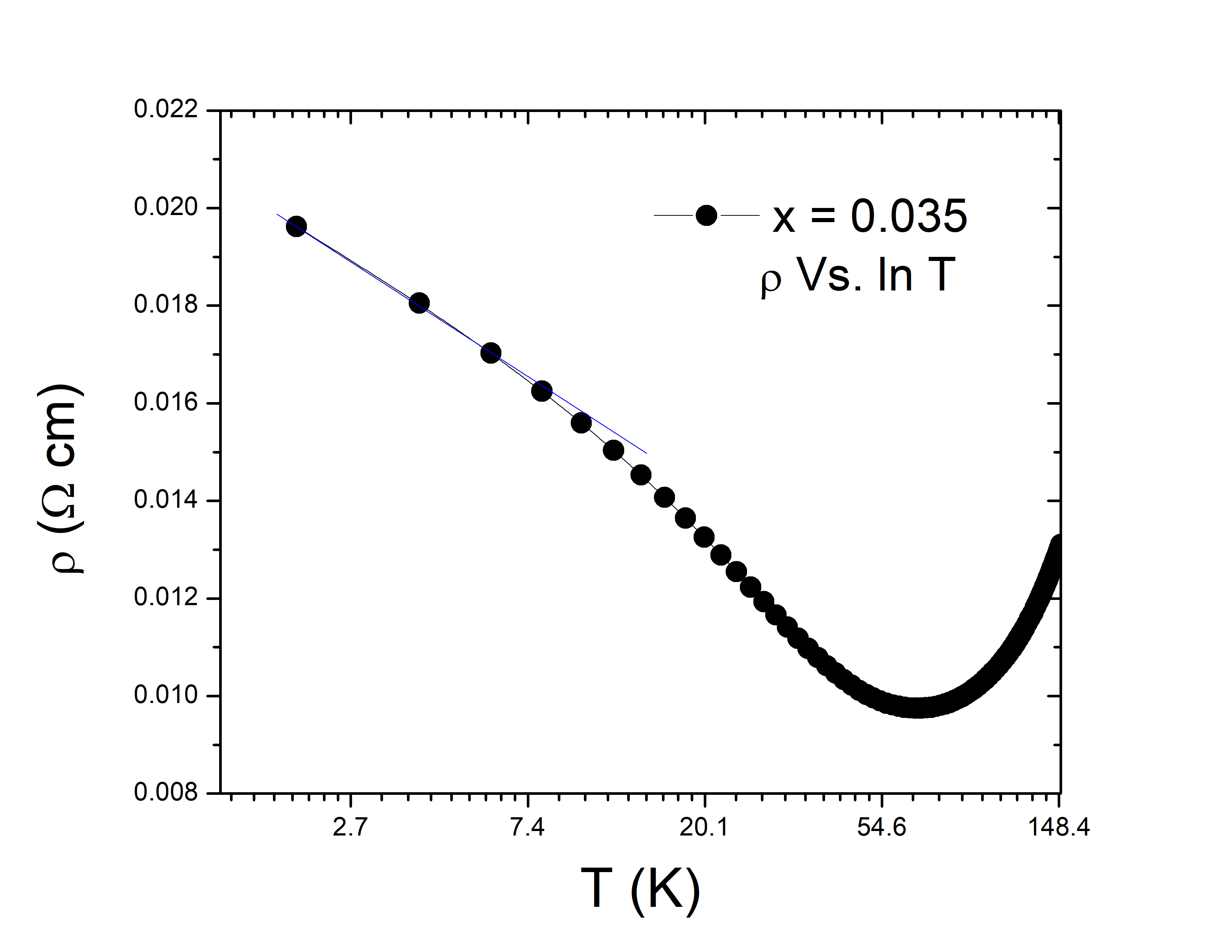}
	\includegraphics[width=0.48\textwidth]{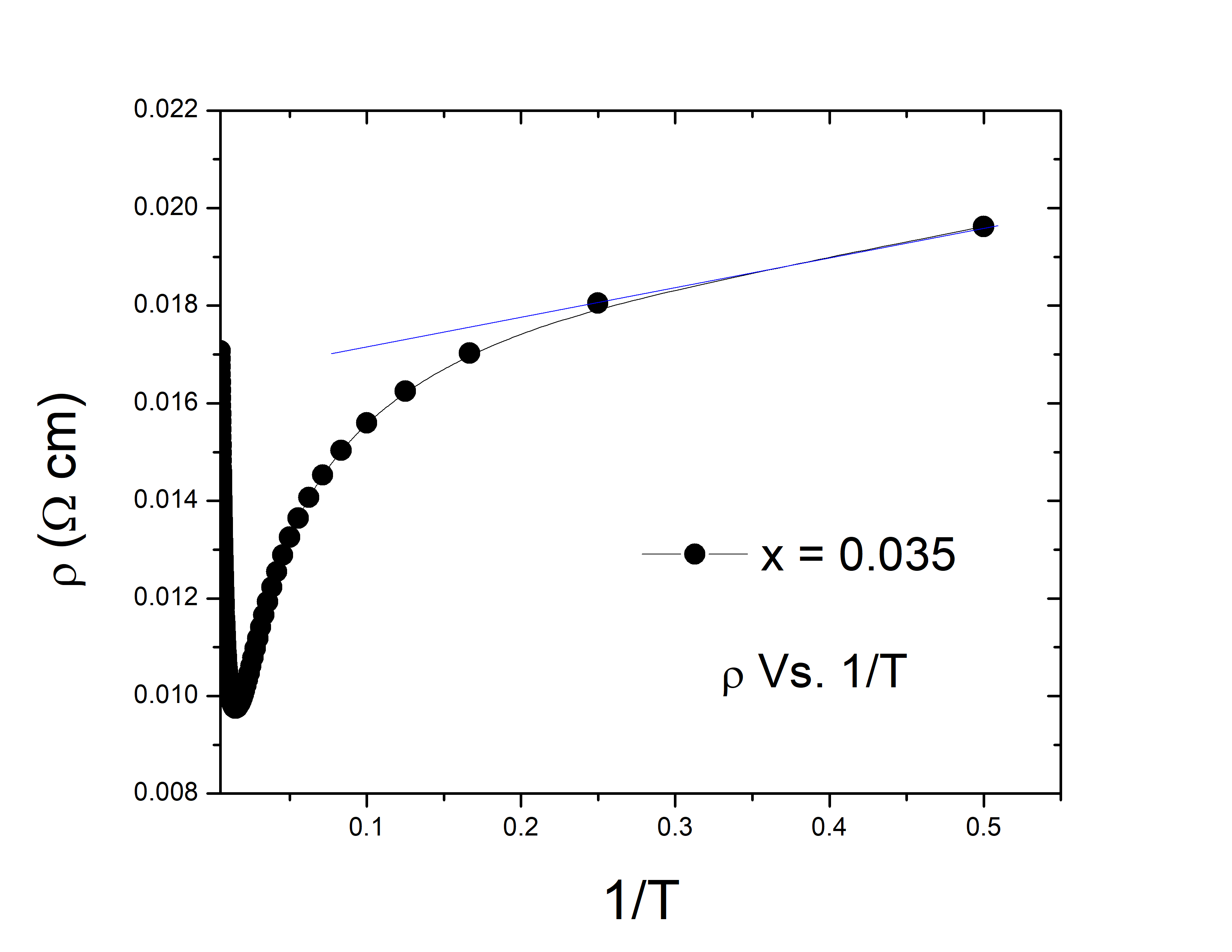}
	\includegraphics[width=0.48\textwidth]{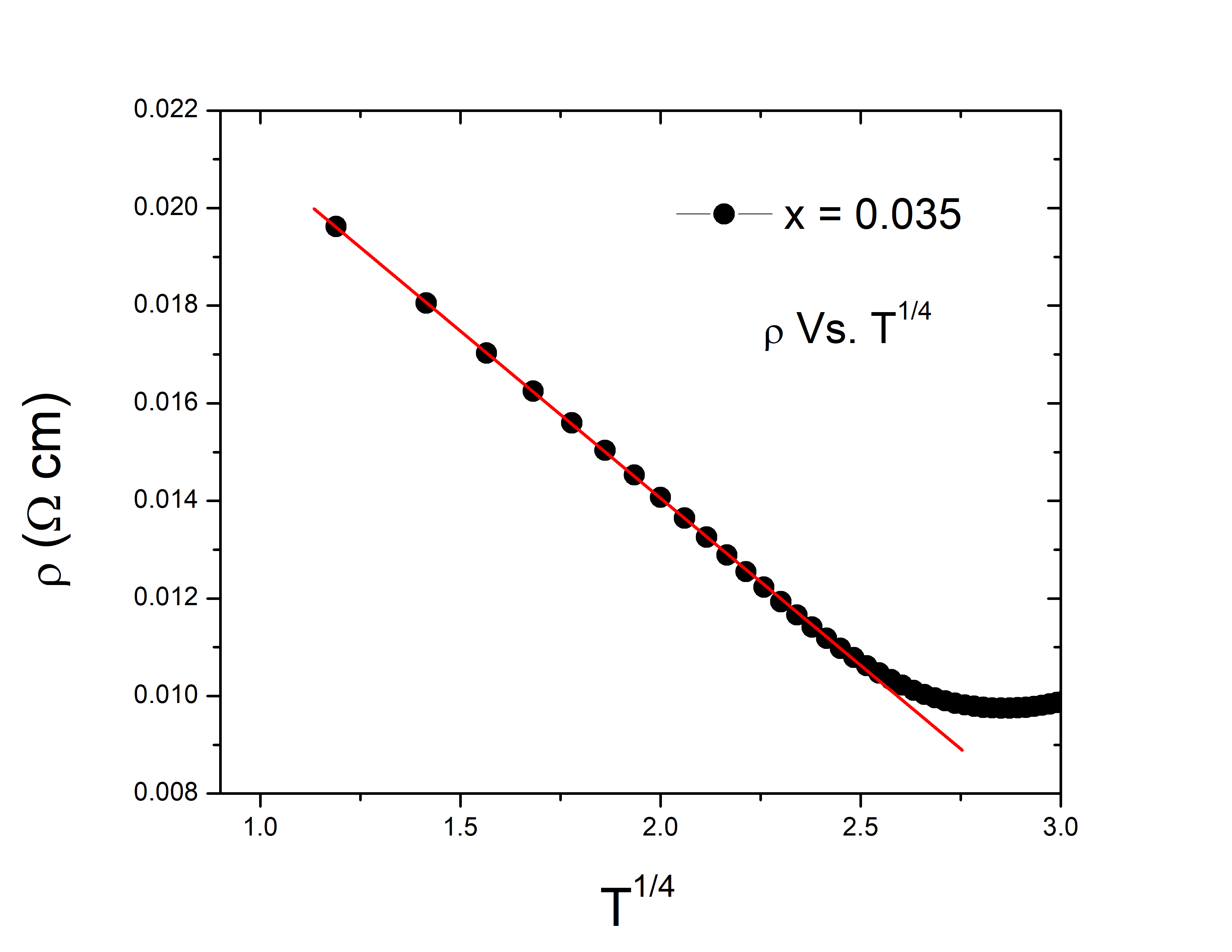}
	\includegraphics[width=0.48\textwidth]{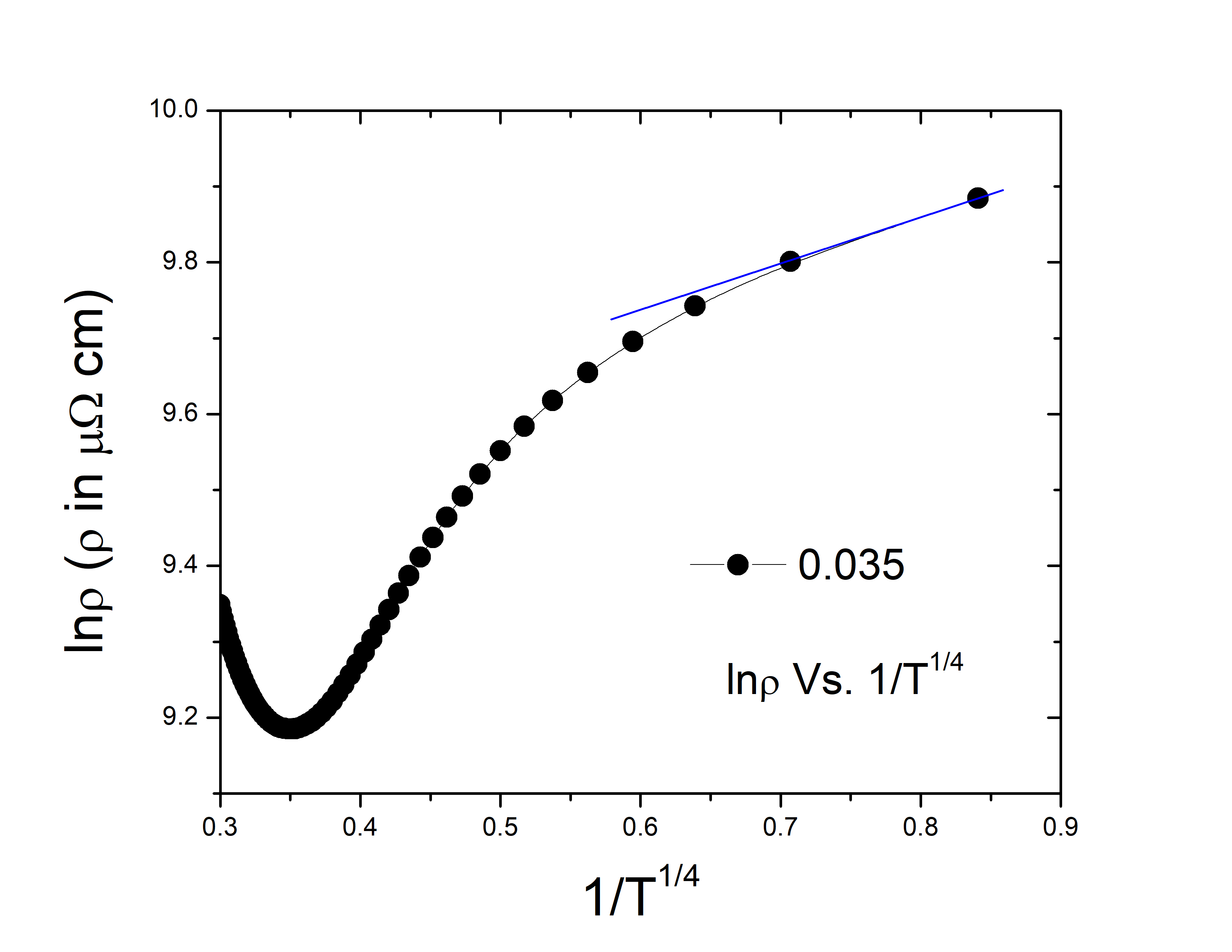}
	%\captionsetup{justification=raggedright,singlelinecheck=false}
	\caption{This figure shows the comparison of various scalings for x = 0.035. Evidently, only in panel (b) the best linear variation is seen; for all other cases the variation is non-linear. Hence, it is reasonable to conclude that the low-temperature resistivity of x = 0.035 sample exhibits a peculiar $\rm -T^{1/4}$ temperature dependence.}
	\label{rho035}
\end{figure*}

\begin{figure*}[!]
	%\centering
	\includegraphics[width=0.48\textwidth]{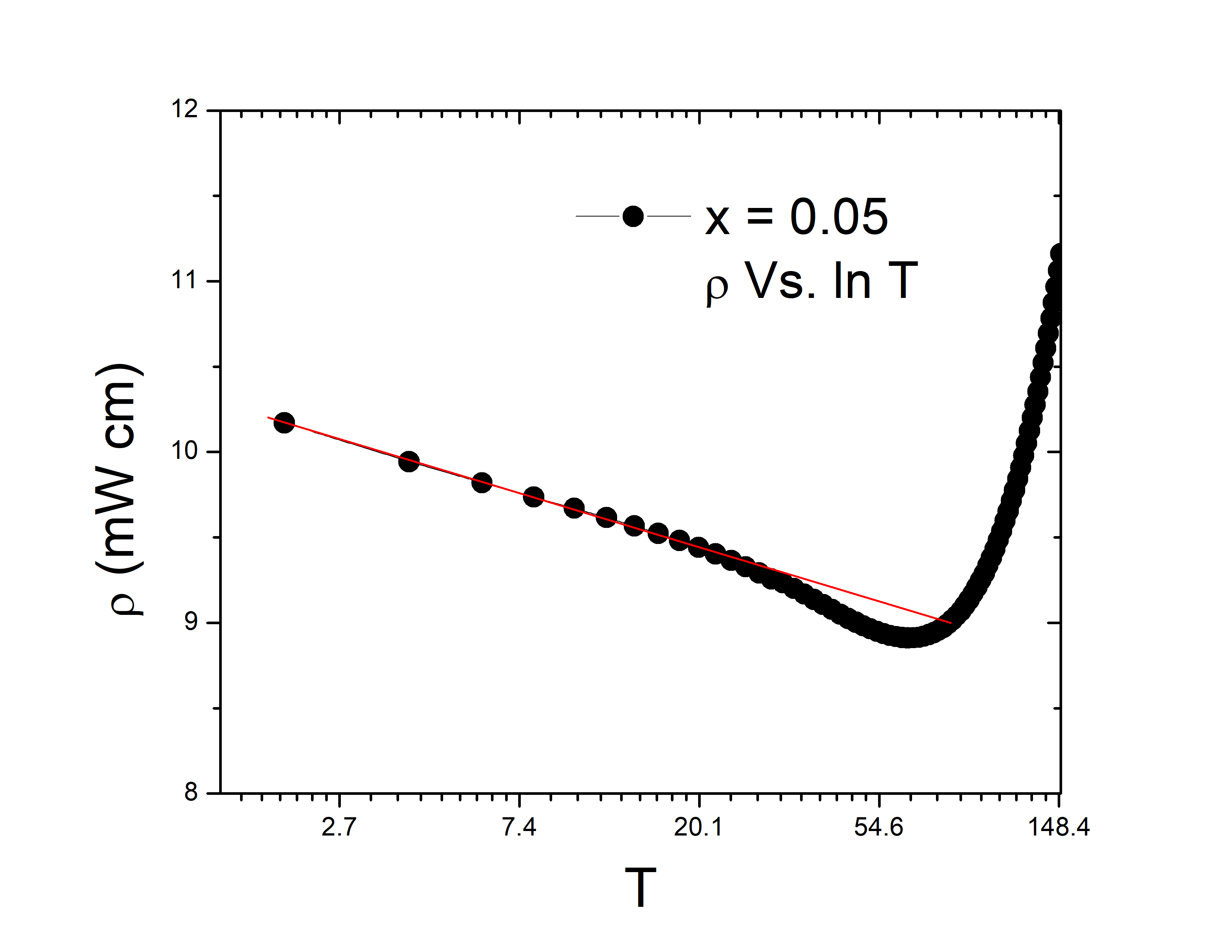}
	\includegraphics[width=0.48\textwidth]{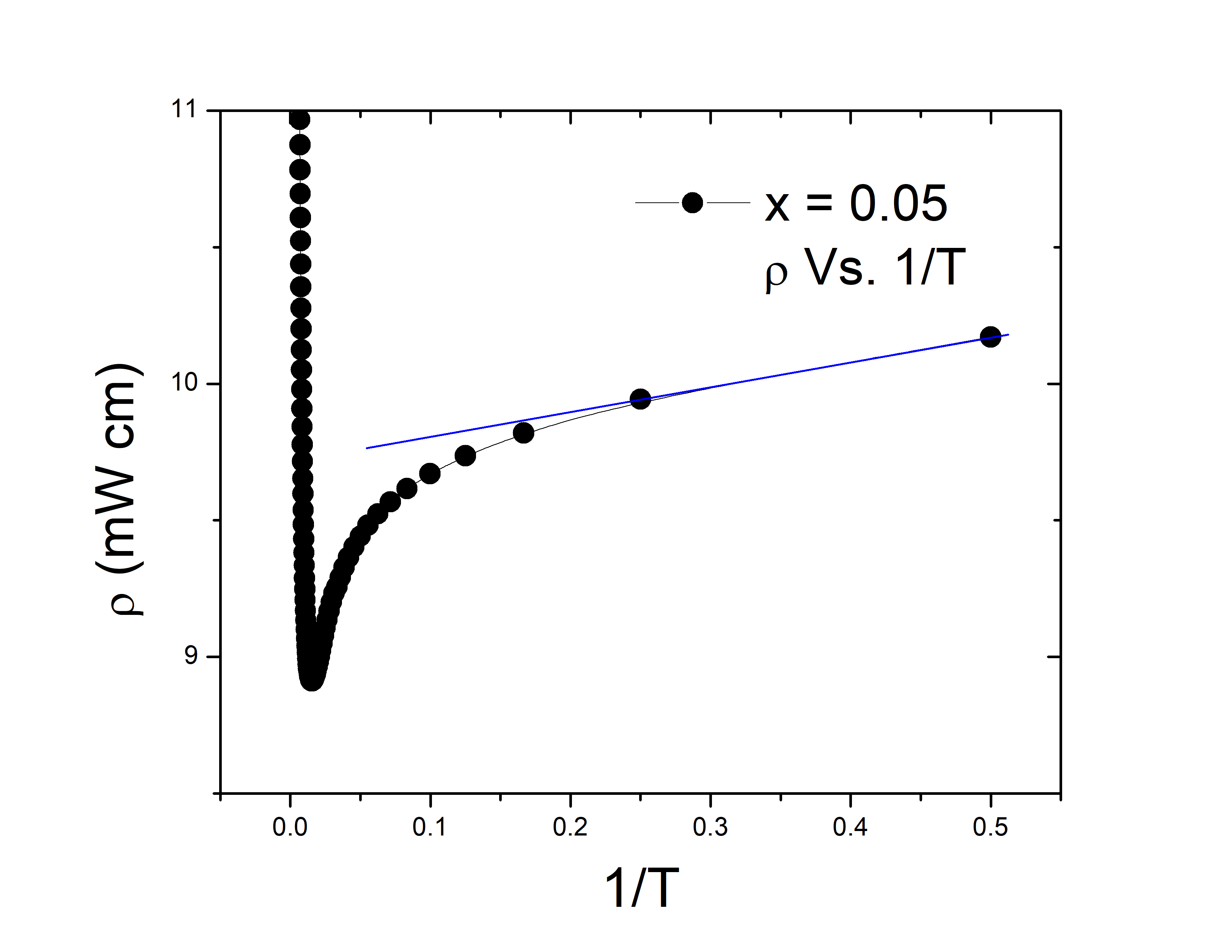}
	\includegraphics[width=0.48\textwidth]{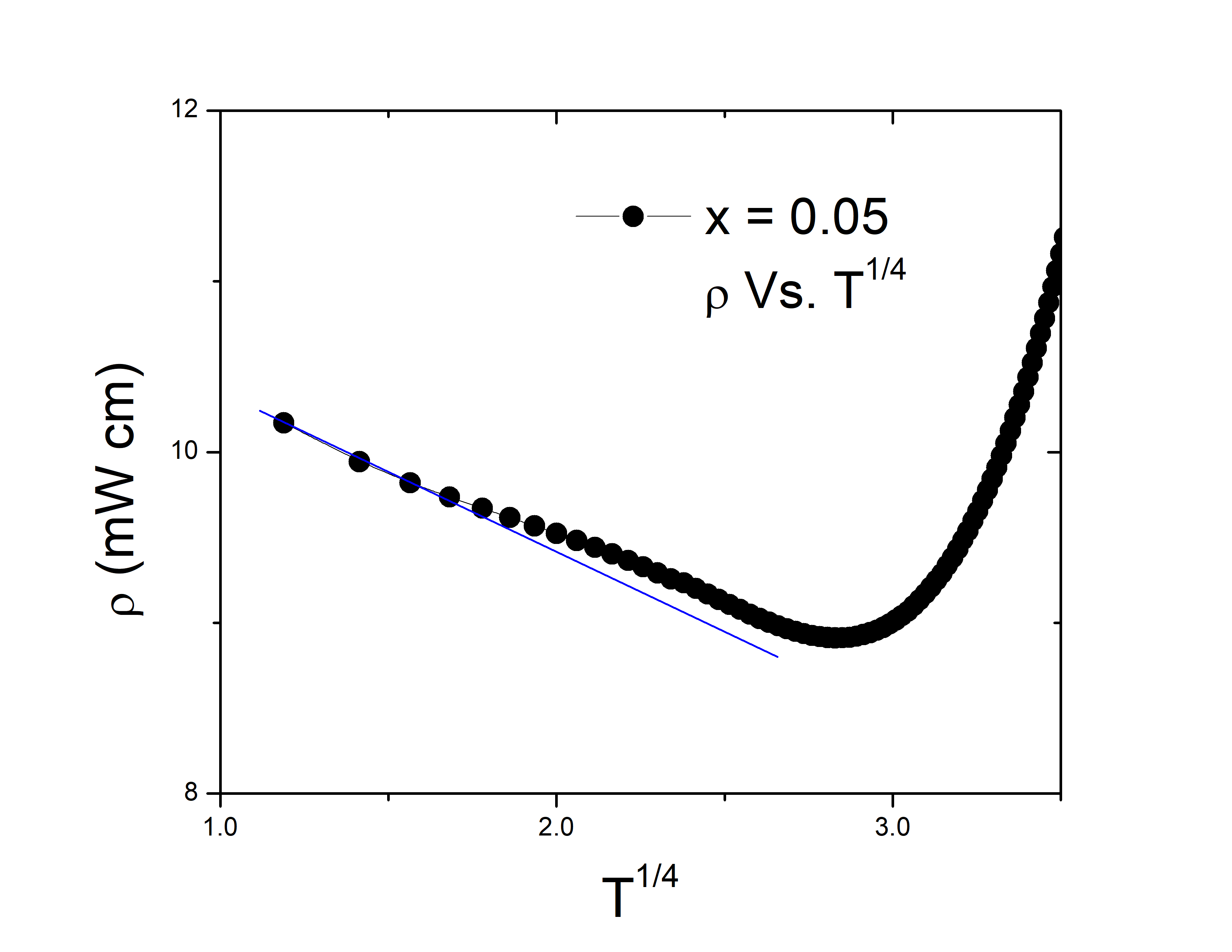}
	\includegraphics[width=0.48\textwidth]{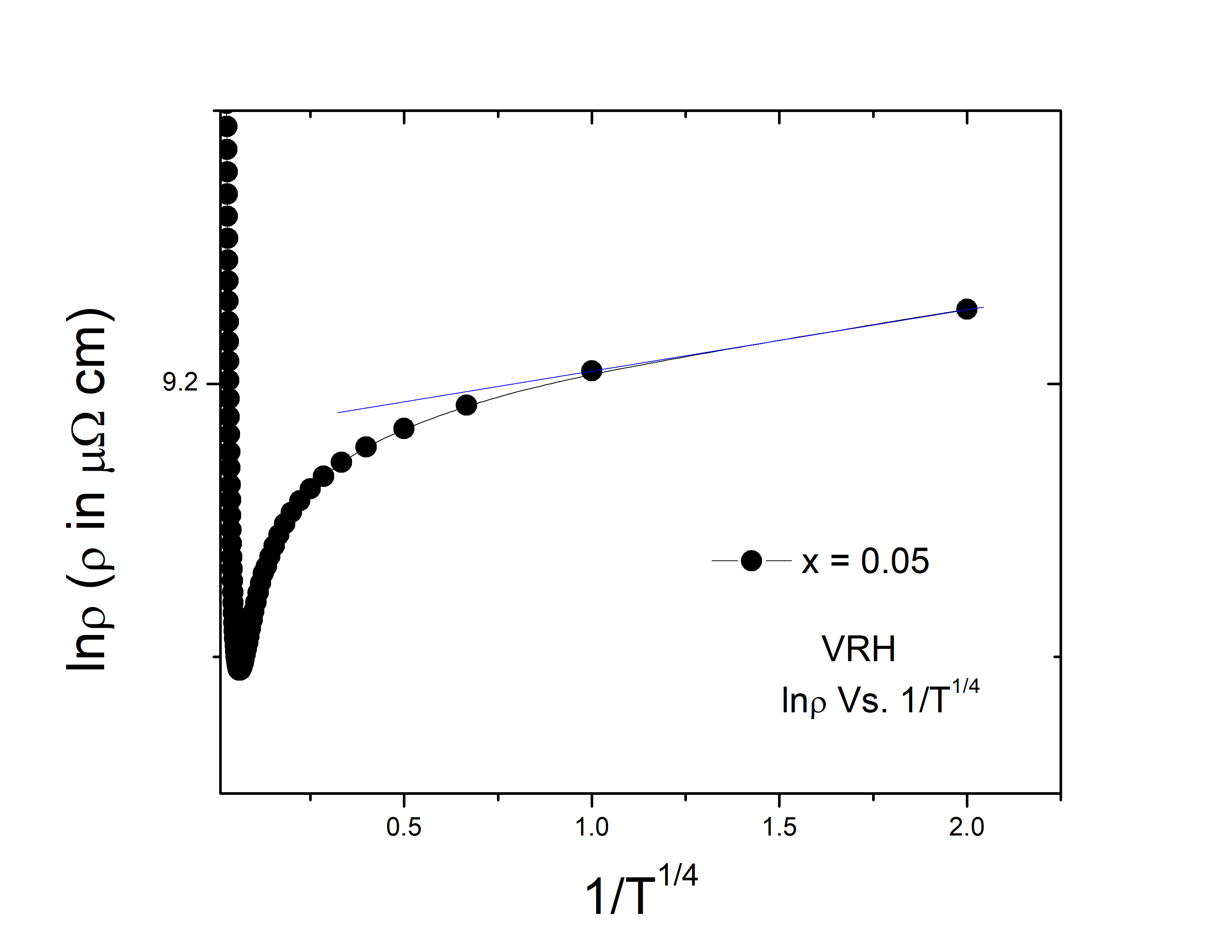}
	%\captionsetup{justification=raggedright,singlelinecheck=false}
	\caption{This figure shows the comparison of various scalings for x = 0.05. Evidently, only in panel (d) the best linear variation is seen; for all other cases the variation is non-linear. Hence, it is reasonable to conclude that the low-temperature resistivity of x = 0.05 sample exhibits a $\rm -lnT$ temperature dependence expected for a QBT scenario.}
	\label{rho05}
\end{figure*}

\begin{figure*}[!]
	%\centering
	\includegraphics[width=0.48\textwidth]{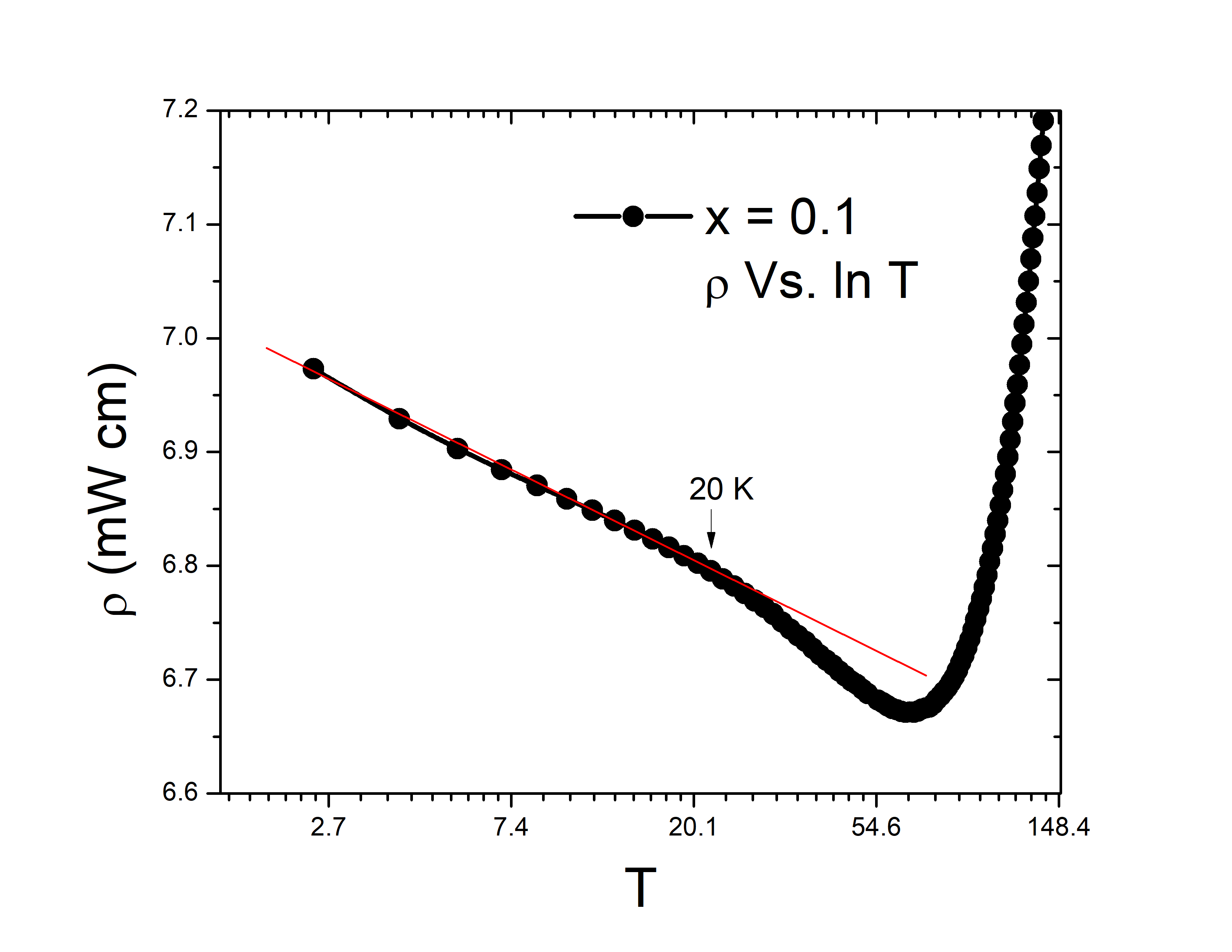}
	\includegraphics[width=0.48\textwidth]{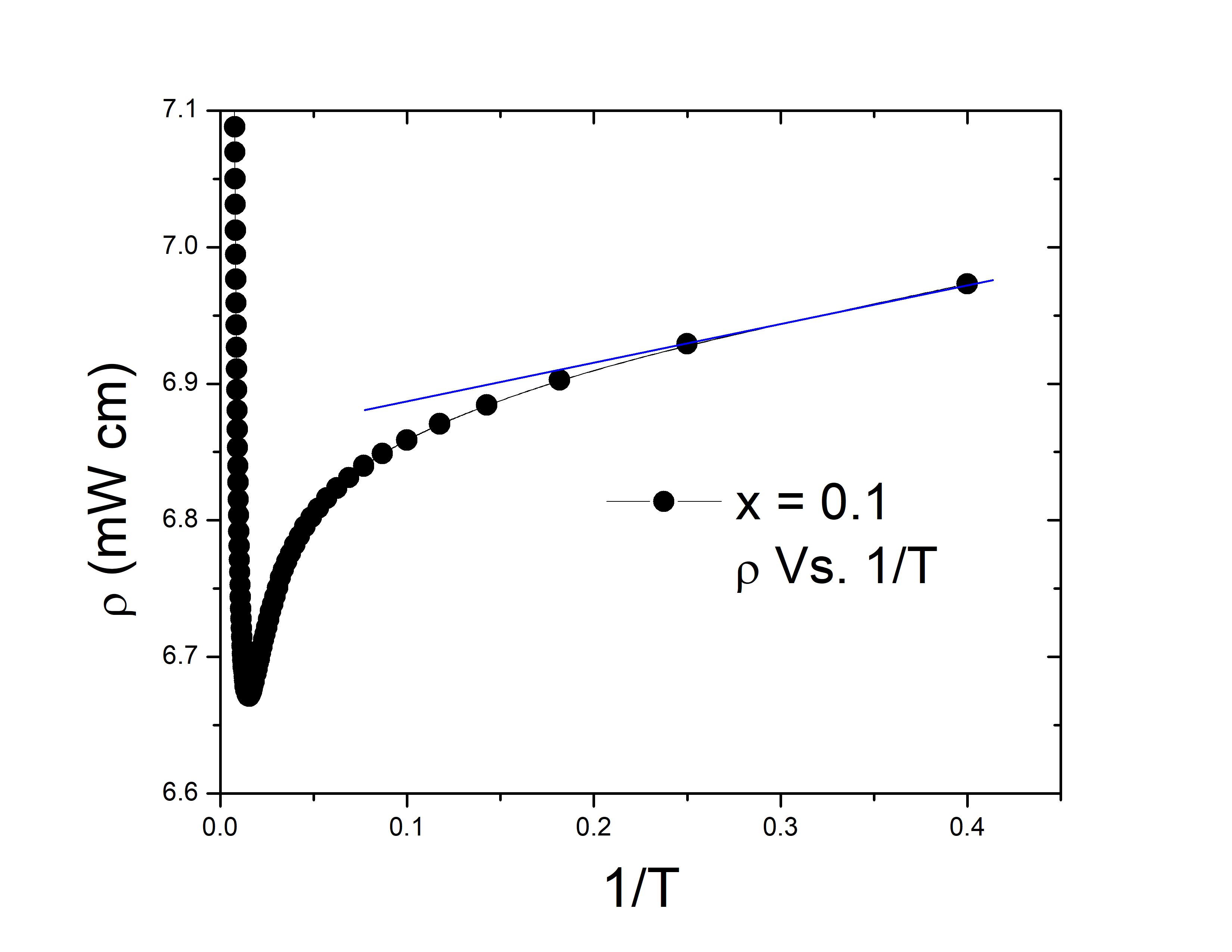}
	\includegraphics[width=0.48\textwidth]{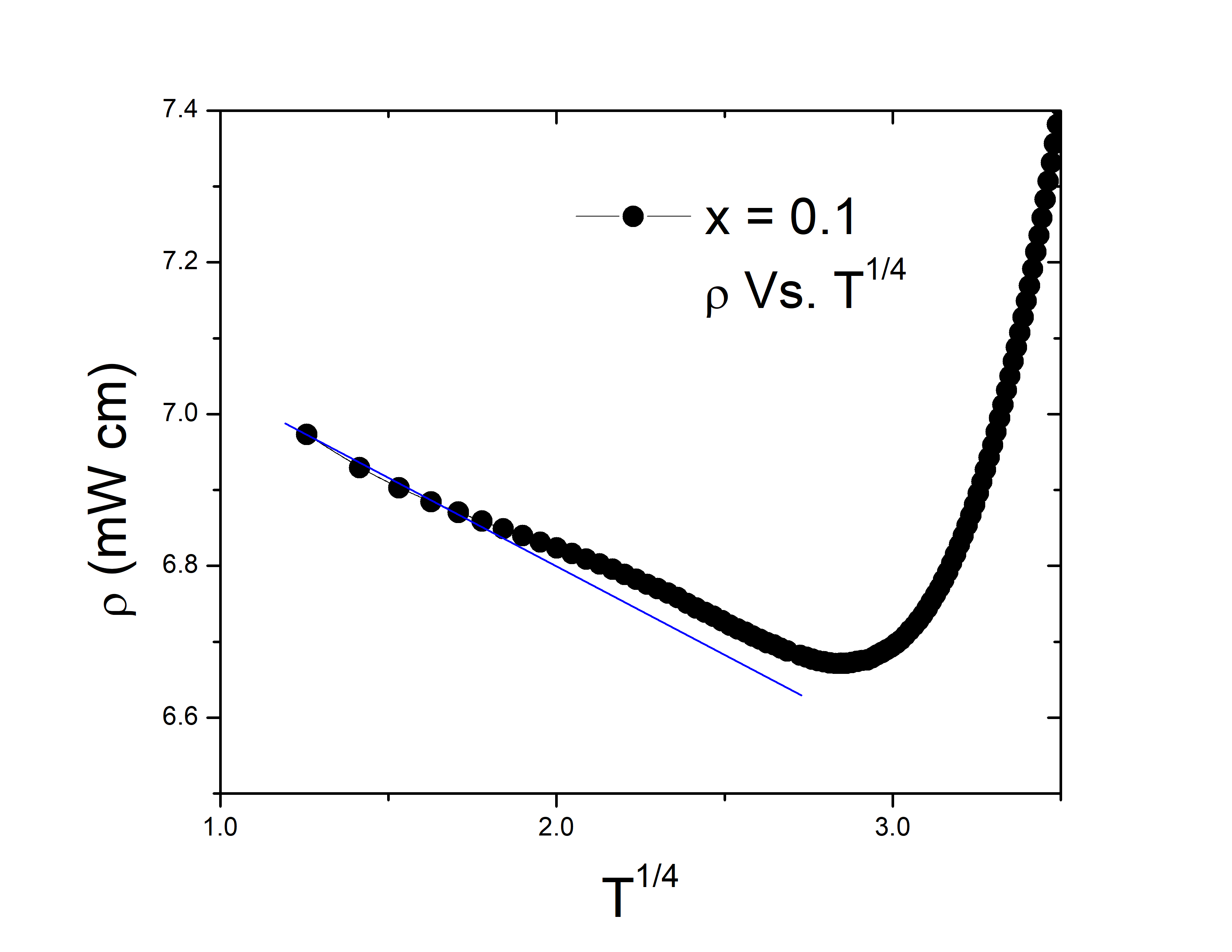}
	\includegraphics[width=0.48\textwidth]{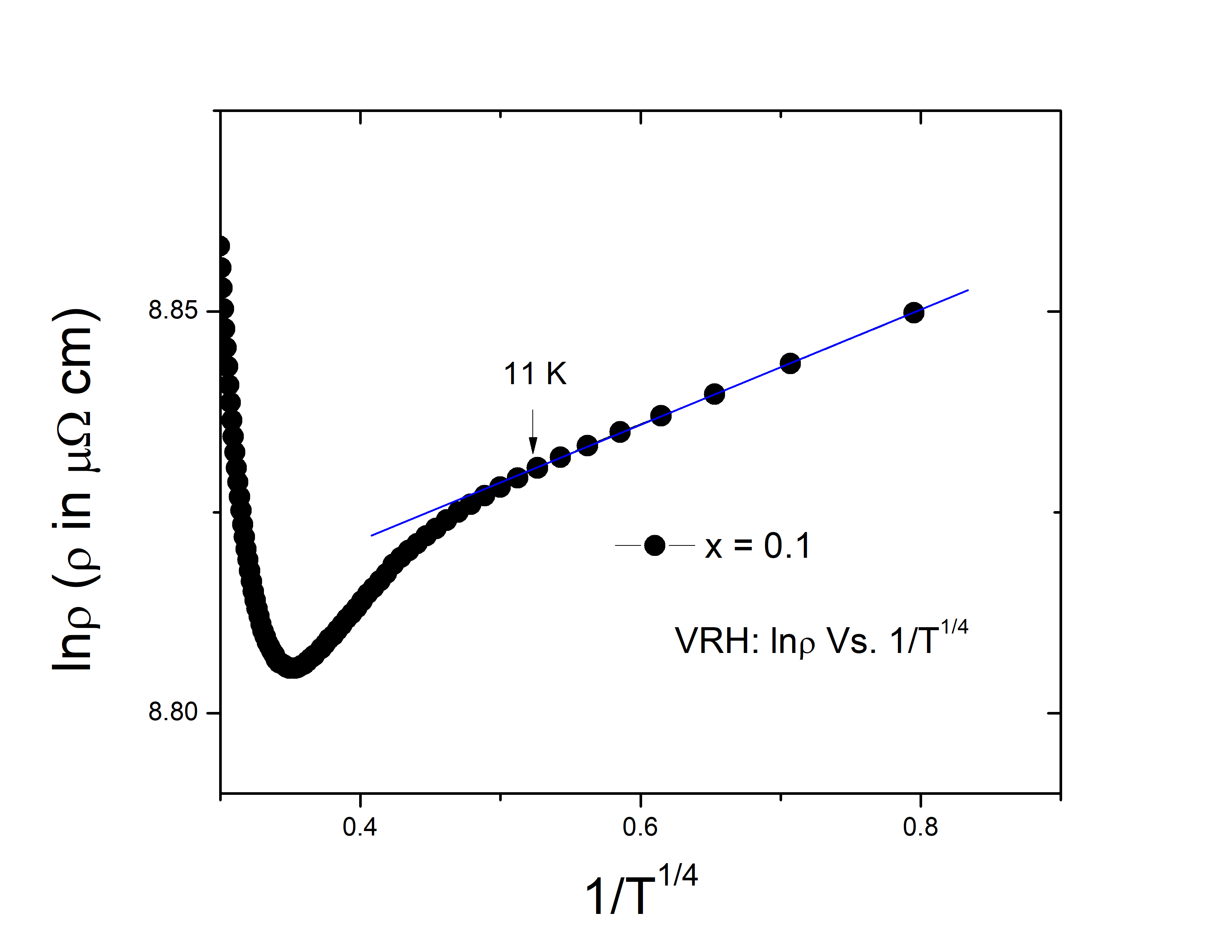}
	%\captionsetup{justification=raggedright,singlelinecheck=false}
	\caption{This figure shows the comparison of various scalings for x = 0.1. Here, the linear behavior can be seen in panels (c) and (d). However, the linearity extends up to about 20 K with $\rm -lnT$ scaling but only till 11 K for the VRH model. For the other two cases the linear variation is not satisfactory. Clearly, as Bi-doping is increasing the behavior is tending towards the variable range hopping which is a characteristic of the disordered systems. However, the influence of QBT on charge transport is still manifested as -lnT temperature dependence extends up to 20 K.}
	\label{rho10}
\end{figure*}

\begin{figure*}[!]
	%\centering
	\includegraphics[width=0.48\textwidth]{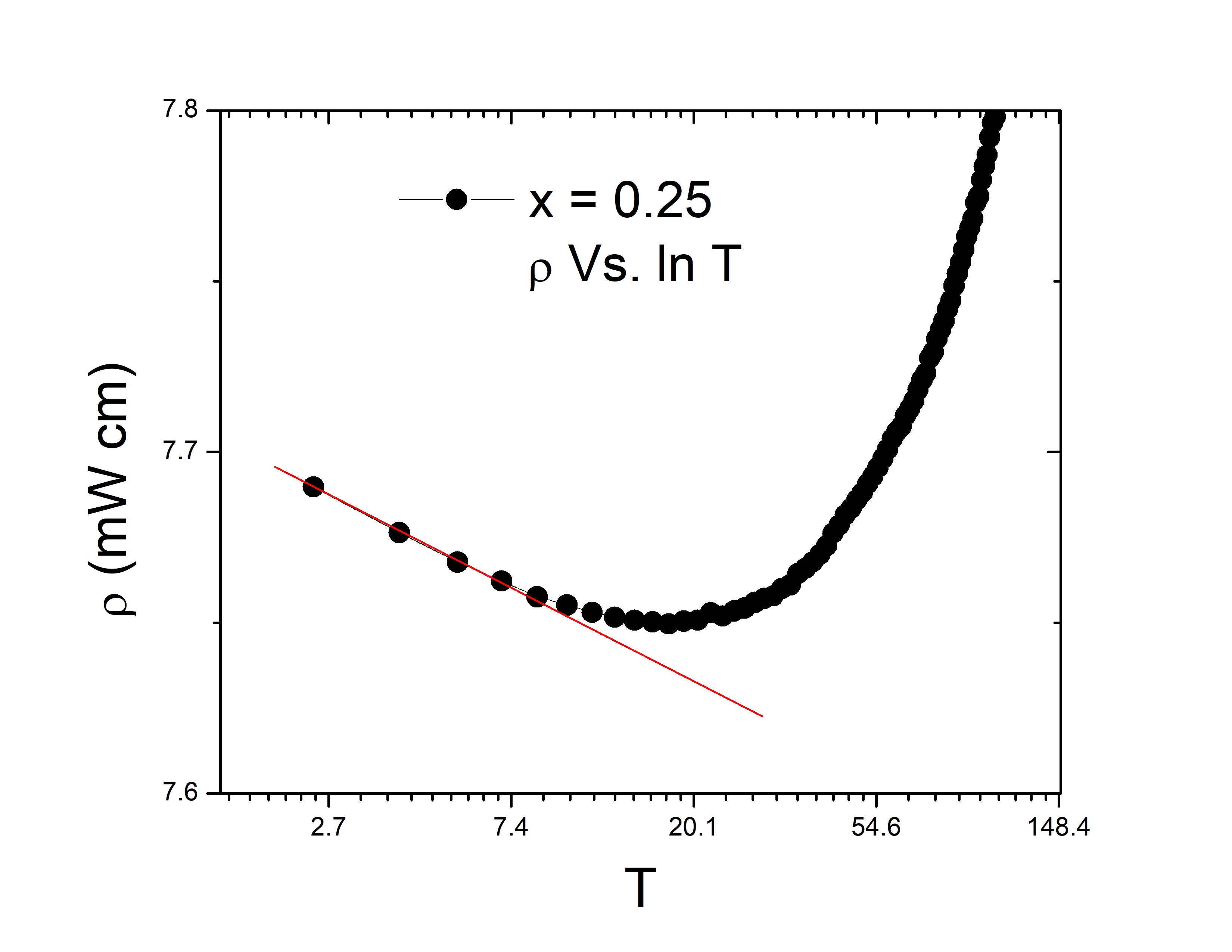}
	\includegraphics[width=0.48\textwidth]{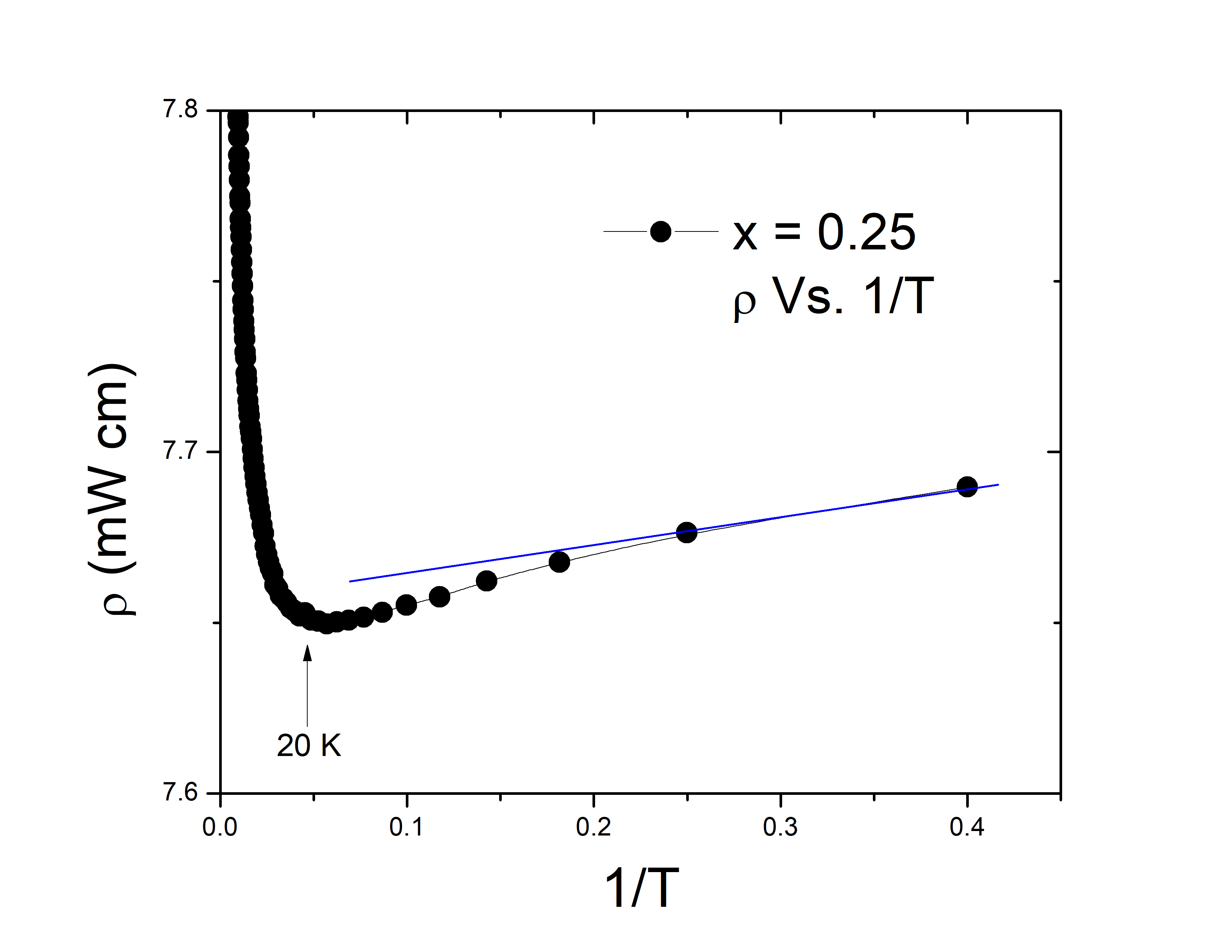}
	\includegraphics[width=0.48\textwidth]{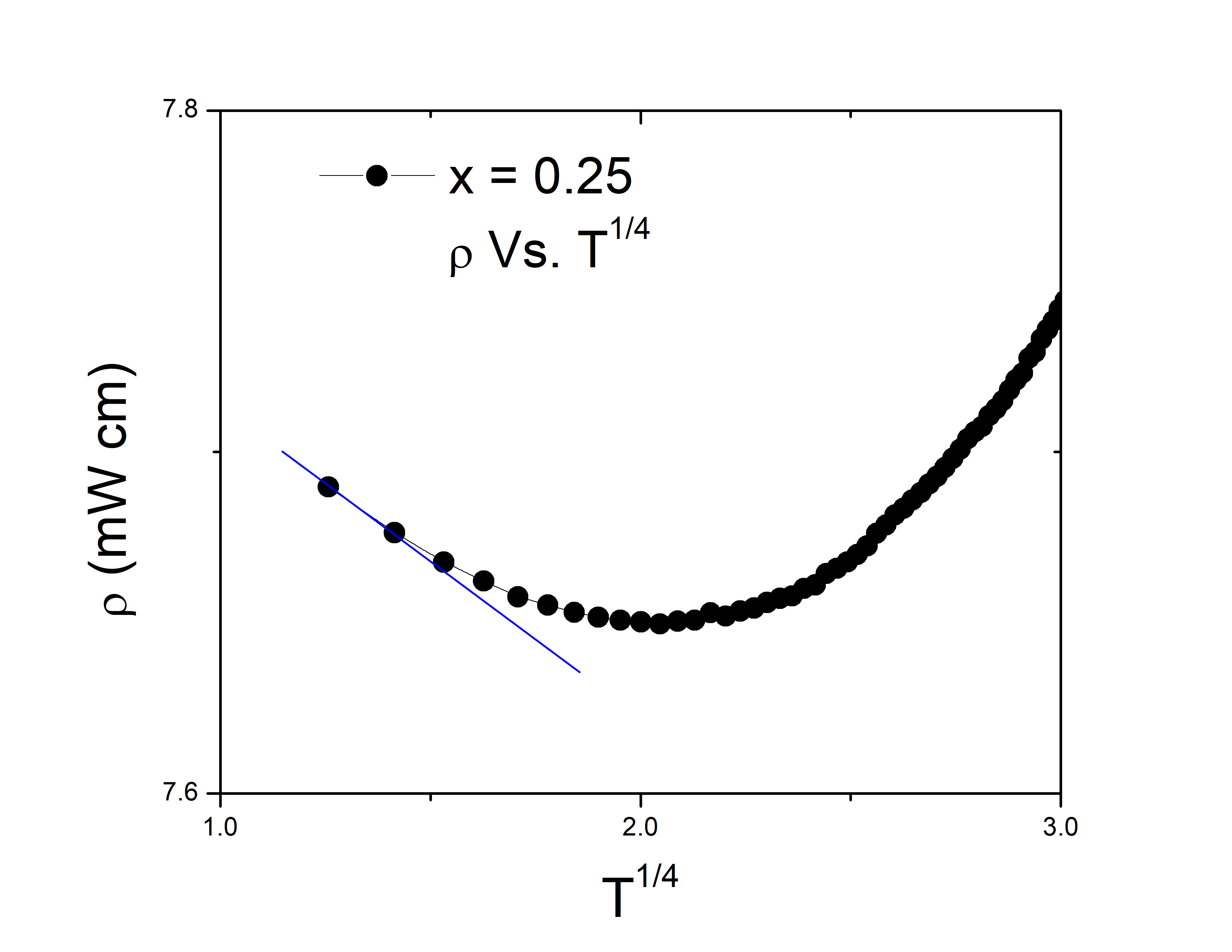}
	\includegraphics[width=0.48\textwidth]{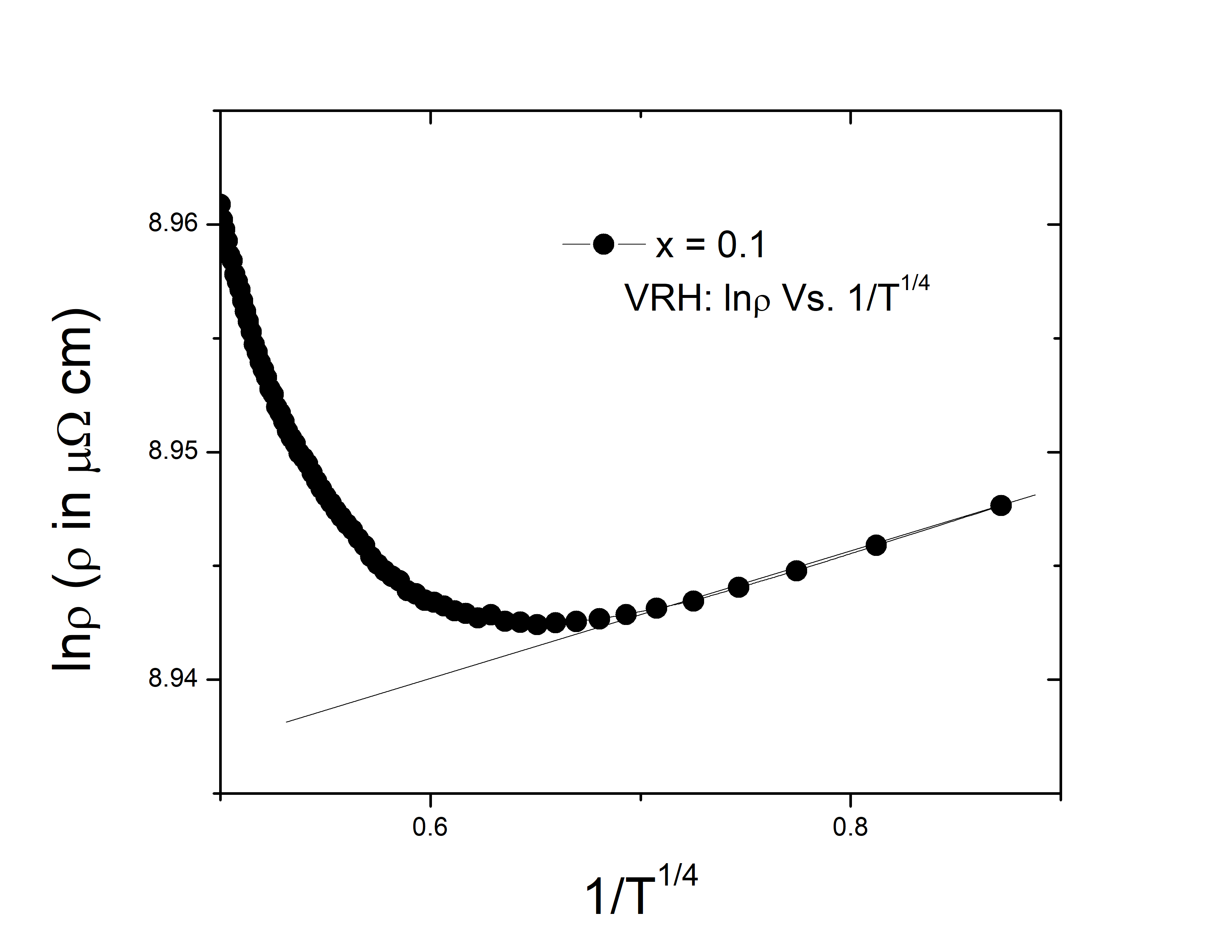}
	%\captionsetup{justification=raggedright,singlelinecheck=false}
	\caption{Figure. S5. This figure shows the comparison of various scalings for x = 0.25. Here, the linear behavior can be seen in panels (c) and (d). However, the linearity is more pronounced in the case of variable range hopping models, which suggests that the $\rm -lnT$ dependence is becoming increasingly suppressed in the presence of increasing Bi disorder. This is also reflected from the magnitude of $\rho$ which has also been slightly enhanced for the 25 \% sample.  In the other two cases the linear variation is simply not satisfactory.}
	\label{rho25}
\end{figure*}

\begin{sidewaystable}
	\vspace{10cm}
	\caption{\textbf{The lattice parameter, O(48f), and various bond distances: Sm-O, Bi-O, Sm-O’, Bi-O’, Ir-O, Ir-O’, Ir-Ir, and bond angle Ir-O-Ir are tabulated for various x}}
	\begin{tabular}{>{\centering\arraybackslash}m{3cm}|>{\centering\arraybackslash}m{2cm}|>{\centering\arraybackslash}m{2cm}|>{\centering\arraybackslash}m{2cm}|>{\centering\arraybackslash}m{2cm}|>{\centering\arraybackslash}m{2cm}|>{\centering\arraybackslash}m{2cm}|>{\centering\arraybackslash}m{2cm}|>{\centering\arraybackslash}m{2cm}}
		\hline
		\hline
		Sm$_{2-2x}$Bi$_{2x}$Ir$_2$O$_7$ & Lattice parameter$a$ & O(48f)($u$) & Sm/Bi-O & Sm/Bi-O' & Ir-O & Ir-O' & $\theta_{Ir-O-Ir}$ & Ir-Ir \\ 
		\hline
		\hline 
		\vspace{0.5cm}
		$x=0$ & 10.323357(20) & 0.33683(2) & 2.483493(3) & 2.235065(3) & 2.033182(2) & 4.279823(7) & 128.6809(3) & 3.649846(5)\\ 
		\hline 
		\vspace{0.5cm}
		$x=0.02 $ & 10.314557(14) & 0.33250(4) & 2.511891(2) & 2.2331669(1) & 2.0121648(1) & 4.276189(5) & 129.96397(1) & 3.646746(4)\\
		\hline 
		\vspace{0.5cm}
		$x=0.035$ & 10.313125(17) & 0.032512(7) & 2.56448(13) & 2.23285 (11) & 1.98113(5) & 4.27559(7) &  133.955(7) & 3.64623 (9)\\ 
		\hline 
		\vspace{0.5cm}
		$x=0.05$ & 10.312731(21) & 0.32433(8) & 2.57012(11) & 2.23277(12) & 1.97765(9) & 4.27542 (8) & 134.389(6) & 3.64609 (9)\\
		\hline 
		\vspace{0.5cm}
		$x=0.1$ & 10.311307(7)  & 0.33296(3) & 2.5078399(1) & 2.2324638(9) & 2.0135417(8) & 4.274843(2) & 129.71936(2) & 3.6455979(1)\\
		\hline 
		\vspace{0.5cm}
		$x=0.25$ & 10.314523(3) & 0.33271(4) & 2.5103934(4) & 2.2331598(4) & 2.0130754(4) & 4.2761755(1) &  129.85222(4)  & 3.6467345(8)\\
		\hline 
		\vspace{0.5cm}
		$x=0.5$ & 10.3176651(20) & 0.33180(3) & 2.5176210(3) & 2.2338400(3) & 2.0089277(2) & 4.2774777(7) & 130.33713(3) & 3.6478455(5)\\
		\hline 
		\vspace{0.5cm}
		$x=1$ & 10.324985(9) & 0.32851(6) & 2.54295 (4) & 2.23543 (3) & 1.99713(9) & 4.28052 (2) & 132.1060(8) & 3.65044 (4)\\
		\hline \hline 
	\end{tabular}
\end{sidewaystable}

\end{document}